\documentstyle[11pt]{article}

\def\be{\begin{eqnarray}}
\def\ee{\end{eqnarray}}
\def\Tr{{\rm Tr}\;}
\def\beitem{\begin{itemize}}
\def\eitem{\end{itemize}}
\def\ben{\begin{enumerate}}
\def\een{\end{enumerate}}
\def\fm{{\;\rm fm}}
\def\roughly#1{\mathrel{\raise.3ex\hbox{$#1$\kern-.75em%
    \lower1ex\hbox{$\sim$}}}}
\def\lsim{\roughly<}

\newcommand{\beq}{\begin{eqnarray}}
\newcommand{\eeq}{\end{eqnarray}}

\def\bi{\begin{itemize}}
\def\ei{\end{itemize}}
\def\ba{\begin{array}}
\def\ea{\end{array}}
\def\ie{{\it i.e}}
\def\eg{{\it e.g.}}

\def\del{\partial}
\def\L{{\cal L}}
\def\M{{\cal M}}
\def\mudr{\mu_{{\rm d.r.}}}
\def\O{{\cal O}}

\long\def\beginomit#1\endomit{}
\def\prl{Phys. Rev. Lett.}
\def\np{Nucl. Phys.}
\def\pl{Phys. Lett.}
\def\pr {Phys. Rev.}

\begin{document}


\begin{titlepage}

\hfill{SNUTP-94-50}

\hfill{hep-ph/9406311}

\hfill{June 1994}

\vskip 0.4in
\begin{center}
{\LARGE\bf An Effective Chiral Lagrangian Approach\\
to Kaon-Nuclear Interactions:\\
\mbox{ } \\
{\it Kaonic atom and kaon condensation}}
\vskip 0.8in
{\large  Chang-Hwan Lee$^a$,  G. E. Brown$^b$, Dong-Pil Min$^a$ and Mannque Rho$^c$}\\
{\large a) \it Department of Physics and Center for Theoretical Physics,}\\
{\large \it Seoul National University, Seoul 151-742, Korea}\\
{\large b) \it Department of Physics, State University of New York,} \\
{\large \it Stony Brook, N.Y. 11794, USA. }\\
{\large c) \it Service de Physique Th\'{e}orique, CEA  Saclay}\\
{\large\it 91191 Gif-sur-Yvette Cedex, France}\\
\vskip 0.4in

{\bf ABSTRACT}\\ \vskip 0.1in
\begin{quotation}
On-shell kaon-nucleon scattering, kaonic atom and kaon condensation are
treated on the same footing by means of a chiral perturbation expansion to
the next-to-next-to-leading order (``N$^2$LO").
Constraining the low-energy constants in the chiral Lagrangian
by on-shell $KN$ scattering lengths
and kaonic atom data, the off-shell s-wave scattering amplitude up to 
one-loop order corresponding to N$^2$LO
and the critical density of kaon condensation up to {\it in-medium}
two-loop order are computed. The effects on kaon-proton scattering 
of the quasi-bound $\Lambda (1405)$ and on kaonic atoms and kaon condensation 
of $\Lambda (1405)$-proton-hole excitations through four-Fermi interactions 
are studied to all orders in density within the 
{\it in-medium} two-loop approximation. It is found 
that the four-Fermi interaction terms in the chiral Lagrangian play an
essential role in providing attraction for kaonic atoms, thereby 
inducing condensation but the critical density is remarkably insensitive to
the strength of the four-Fermi interaction that figures in kaonic atoms.
The prediction for the critical density is extremely robust and gives --
for ``natural" values of the four-Fermi interactions --  a rather
low critical density, $\rho_c \lsim 4 \rho_0$. When the BR scaling is
suitably implemented, the condensation sets in at $\rho_c\simeq 2\, \rho_0$
with loop corrections and four-Fermi interactions playing a minor role.
\end{quotation}
\end{center}\end{titlepage}


\section{Introduction}
\indent

In a series of recent short papers\cite{BLRT,LJMR,LBR}, we have discussed 
kaon-nucleon and 
kaon-nuclear interactions in terms of a chiral perturbation expansion with the
objective to predict within the framework of chiral effective Lagrangians
the onset of kaon condensation in dense hadronic matter relevant to 
compact stars that are formed from the gravitational collapse of massive stars. 
This research was
given a stronger impetus by a recent suggestion of 
Brown and Bethe \cite{brownbethe} that if kaon
condensates develop at a matter density $\rho\lsim 4\ \rho_0$
(where $\rho_0\approx 0.16/fm^3$ is normal nuclear matter density)
in the collapse of large
stars, then low-mass black holes are highly likely to form in place of 
neutron stars of the mass greater than 1.5 times the solar mass $M_\odot$.
The purpose of this paper is to provide the details of our previous 
publications, such as the approximations made, inherent uncertainties involved
etc. with complete and explicit formulae
that we were unable to supply in the letter papers because of the space 
limitation. In addition, some additional justifications that have in 
the meantime been uncovered and understood better will be discussed.

Ever since the first paper of Kaplan and Nelson\cite{KN},
there have been numerous investigations on
kaon condensation in dense neutron-star matter as well as in nuclear matter
based both on effective chiral
Lagrangians\cite{BLRT,BKR,PW,BKRT,TPL} and on phenomenological
off-shell meson-nucleon interactions\cite{YNMK,lutz}. The two ways of 
addressing the problem gave conflicting results with 
the chiral Lagrangian approaches generally
predicting a relatively
low critical density, $\rho_c\sim (2-4) \rho_0$, while the phenomenological
approaches based more or less on experimental inputs giving results that
tend to exclude condensation at a low enough density to make it relevant
in the collapse process.

We should stress that it is not our purpose here 
to clarify what makes the phenomenological approach
differ from the chiral Lagrangian approach. Instead, 
our objective in this paper is to
do as consistent and systematic a calculation as possible in the context of
chiral perturbation theory. This effort is largely motivated from the 
strong-interaction point of view by the recent success in confronting,
in terms of chiral dynamics, the classic nuclear physics problems such as
nuclear forces\cite{wein,vankolck} and exchange currents\cite{mr,PMR,PTK}. 
This paper addresses the problem of applying chiral perturbation theory 
to multi-hadron
systems that contain the strange-quark flavor. While the standard problems
of nuclear physics such as nuclear forces and exchange currents involve the 
chiral quarks $u$ and $d$ for which the
mass scale involved is small compared with the typical QCD chiral symmetry
breaking scale $\Lambda_\chi \sim 1$ GeV
so that a simultaneous expansion in derivatives
and quark mass matrix is justified, here the strange-quark mass which is not
small renders the expansion in the quark mass a lot more delicate and hence
the low-order expansion highly problematic. This caveat has to be kept
in mind in assessing the validity of the procedure we will adopt.

Another problem of potential importance is that both kaonic atom and kaon
condensation introduce an additional scale, namely the matter density $\rho$ or
more precisely the Fermi momentum $k_F$. So far in low-order chiral 
perturbation calculations, the result depended on the order of density
dependence included in the calculation. In fact, 
{one of the important differences} 
between the chiral Lagrangian approach and the phenomenological
model approach arose at the order $\rho^2$.  It is thus clear that
one has to be consistent in the chiral counting, not only with respect to the
usual expansion parameters practiced in free space,
but also with respect to the density expansion.
This then raises the question of how to treat the dynamics involved in
the non-strange sector as well as in the strange sector. So far the dynamics
in the non-strange sector is assumed to be given by what we know from 
nuclear phenomenology that is mostly given in terms of meson-theoretic
approaches combined with many-body techniques, 
and perturbations in the strange direction are treated 
in terms of chiral Lagrangian at
tree order or at most one-loop order. The problem with this is that
there is no consistency between the two sectors as regards chiral
symmetry and other constraints of QCD. Indeed so far
nobody has been able to describe correctly nuclear ground state
(including nuclear matter) starting from chiral Lagrangians, so one 
is justified to wonder how a low-order chiral Lagrangian calculation of
kaon condensation without regard to the normal matter can be trusted.
We cannot offer a solution to this problem here but we will make 
an effort to point out the salient points that are closely related
to this issue.

Following the recent development of nuclear chiral dynamics
\cite{wein,vankolck,mr,PMR,PTK}, we incorporate
spontaneously broken chiral symmetry using
the Jenkins-Manohar heavy-baryon chiral Lagrangian
\cite{JM} as extended in \cite{LJMR} to $\O (Q^3)$ to describe
s-wave kaon nucleon scattering to one loop order in chiral perturbation theory
(ChPT). In addition to the usual octet and decuplet baryons and the octet
pseudo-Goldstone fields, the $\Lambda (1405)$ was found to figure importantly
in the kaon-nucleon process. This is because as is well-known, the
$\Lambda (1405)$ (which we denote $\Lambda^\star$ for short) 
influences strongly the
amplitude of the $K^- p$ scattering near threshold and hence kaon-nuclear 
interactions in kaonic atom \cite{kaonicatom}
and kaon condensation involving protons as in 
``nuclear stars." We introduce this state as an elementary field as discussed
in \cite{LJMR}. The reason for this is that first of all, the $\Lambda^\star$ is
a bound state and hence cannot be described by a finite chiral perturbation
expansion and secondly in the Callan-Klebanov skyrmion description \cite{SMNR},
it is a configuration of a $K^-$ wrapped by an $SU(2)$ soliton and hence is as
``elementary" as the even-parity $\Lambda (1115)$ of the octet baryon.

In addition to these terms operating in the single-baryon sector, we need
terms that involve multi-baryon fields in the Lagrangian for describing
many-body systems. There have been discussions of four-Fermi interaction
terms in non-strange sectors \cite{wein,vankolck,PMR,MPW}. We find that in 
the s-wave kaon-nuclear sector, two such
four-Fermi interaction terms involving $\Lambda^\star$ can intervene. 
In p-wave kaon-nuclear interactions, there can 
be more four-Fermi interactions as they can involve
the entire battery of the octet and decuplet but we will not be concerned with 
them in this paper.

By a straightforward extension of an amplitude whose parameters are fixed
by on-shell kaon-nucleon scattering, 
we are able to almost (but not quite) uniquely
predict an off-shell kaon-nucleon amplitude relevant for
kaonic-atom as well as kaon-condensation phenomena.
The predicted off-shell amplitude was found to be in fair agreement
with the phenomenological fit \cite{Steiner}.
This off-shell amplitude provides the kaon self-energy in
linear density approximation, equivalent to the usual optical potential
approximation.  The critical density obtained in this approximation is a bit
higher than that obtained before at tree order but still in the regime quite
relevant to the stellar collapse. If one goes 
beyond the linear density approximation which would be required in a 
simultaneous expansion in all the scales involved, 
four-Fermion interactions come into play. 
For the s-wave kaon condensation process, there are two independent four-Fermi
interactions with arbitrary constants.
In order to fix these parameters,
we appeal to the recent data on kaonic
atoms\cite{kaonicatom}. We are able to fix unambiguously only one of the two 
constants with the presently available data. While the other constant
remains free, its sign and magnitude can, however,
be constrained by a ``naturalness"
condition and furthermore the physical quantities that we are interested in
turn out to be rather insensitive to the free parameter.

The four-Fermi interactions -- which are higher order in density --
play an important role for giving rise to an attraction for kaonic atoms.
This attraction certainly comes in for {\it pushing} the system toward 
condensation. However, they remain ``irrelevant" and become suppressed
at the kinematic regime in which condensation occurs. As a consequence,
their influence on the critical density is quite weak: The strength
of the four-Fermi interactions, which cannot be pinned down precisely
at present, does not figure importantly in the condensation phenomena.

We do not consider six-Fermi and higher-Fermi
interactions as they would be further suppressed by the scale set by
the chiral symmetry breaking scale $\Lambda_\chi\sim 1$ GeV.

The paper is organized as follows. The effective chiral Lagrangian to
${\O (Q^3)}$ in the chiral counting, consisting of the octet pseudo-Goldstone
bosons and the octet and decuplet baryons that figure in our calculation,
is given in section 2. In section 3, we calculate to one-loop order,
corresponding to N$^2$LO, both on-shell
and off-shell KN scattering amplitudes. Some issues regarding
Adler's soft-meson conditions in chiral perturbation theory
are also discussed. Kaonic atom is treated in section 4
and kaon condensation in sections 5 and 6. In section 7, we mention some of
the unsolved open issues in the problem. Detailed formulas are collected
in the appendices.


\section{Effective Chiral Lagrangian}
\indent

We start by writing down the effective chiral Lagrangian that we shall
use in the calculation.
Let the characteristic momentum/energy scale that we are
interested in be denoted $Q$. The standard chiral counting orders 
the physical amplitude as a
power series in $Q$, say, $Q^\nu$, with $\nu$ an integer. To leading order, the
kaon-nucleon amplitude $T^{KN}$ goes as ${\cal O}(Q^1)$, to next order as
${\cal O}(Q^2)$
involving no loops and to next to next order ({\it i.e.}, N$^2$LO)
at which one-loop graphs enter as ${\cal O}(Q^3)$.
Following Jenkins and Manohar \cite{JM}, we denote
the velocity-dependent octet baryon fields $B_v$, the
octet meson fields exp$(i\pi_aT_a/f) \equiv\xi$,
the velocity-dependent decuplet baryon fields $T_v^{\mu}$,
the velocity four-vector $v_{\mu}$ and the
spin operator $S_v^{\mu}$ ($v\cdot S_v=0$, $S_v^2=-3/4$),
the vector current $V_{\mu}=[\xi^{\dagger},\partial_{\mu}\xi]/2$ and the
axial-vector current $A_{\mu}=i\{\xi^{\dagger},\partial_{\mu}\xi\}/2$,
and write the Lagrangian density to order $Q^3$,
relevant for the low-energy s-wave scattering\footnote{The relevant
terms in component fields useful for calculating Feynman diagrams
are given in Appendix A.}, as
   \beq
   {\cal L}^{(1)} &=&
          \Tr{\bar B }_v(iv\cdot{\cal D})B_v
          +2D\Tr{\bar B}_vS_v^{\mu}\{A_{\mu},B_v\}
          +2F\Tr{\bar B}_vS_v^{\mu}[A_{\mu},B_v]\nonumber\\
       && -{\bar T}_v^{\mu}(iv\cdot{\cal D}-\delta_T)T_{v,\mu}
          +{\cal C}({\bar T}_v^{\mu}A_{\mu}B_v+\bar{B}_vA_{\mu}T_v^{\mu})
          + 2{\cal H}\bar{T}_v^{\mu}(S_v\cdot A)T_{v,\mu} \label{l1}\\
   {\cal L}^{(2)} &=&
          a_1\Tr{\bar B}_v\chi_+B_v
          +a_2\Tr{\bar B}_vB_v\chi_+
          +a_3\Tr{\bar B}_vB_v\Tr\chi_+\nonumber\\
       && +d_1\Tr{\bar B}_vA^2B_v+d_2\Tr{\bar B}_v(v\cdot A)^2B_v
          +d_3\Tr{\bar B}_vB_vA^2+d_4\Tr{\bar B}_vB_v(v\cdot A)^2\nonumber\\
       && +d_5\Tr{\bar B}_vB_v \Tr A^2
          +d_6\Tr{\bar B}_vB_v\Tr (v\cdot A)^2
       +d_7\Tr {\bar B}_vA_{\mu}\:\Tr B_vA^{\mu}\nonumber\\
    && +d_8\Tr {\bar B}_v(v\cdot A)\:\Tr B_v (v\cdot A),\label{L2}\\
   {\cal L}^{(3)} &=&
       c_1\Tr {\bar B}_v(iv\cdot{\cal D})^3B_v +g_1\Tr {\bar B}_vA_{\mu}
       (iv\cdot\stackrel{\leftrightarrow}{\cal D})A^{\mu}B_v +g_2\Tr B_vA_{\mu}
       (iv\cdot\stackrel{\leftrightarrow}{\cal D})A^{\mu}{\bar B}_v
       \nonumber\\
    && +g_3\Tr {\bar B}_vv\cdot A(iv\cdot\stackrel{\leftrightarrow}{\cal D})
       v\cdot AB_v
       +g_4\Tr B_vv\cdot A(iv\cdot\stackrel{\leftrightarrow}{\cal D})v\cdot A
       {\bar B}_v\nonumber\\
    && +g_5\left(\Tr  \bar B_v A_{\mu} \Tr (iv\cdot\stackrel{\rightarrow}
       {\cal D}) A^{\mu}B_v
       -\Tr \bar B_v A_{\mu}(iv\cdot \stackrel{\leftarrow}{\cal D})
       \Tr A^{\mu}B_v \right)\nonumber\\
    && +g_6\left(\Tr \bar B_v v\cdot A \Tr B_v (iv\cdot\stackrel{\rightarrow}
       {\cal  D})v\cdot  A
       -\Tr \bar B_vv\cdot A (iv\cdot\stackrel{\leftarrow} {\cal D}
       )v\cdot A \Tr B_v v\cdot A\right)\nonumber\\
    && +g_7 \Tr \bar B_v [v\cdot A,[iD^\mu,A_\mu]]B_v
       +g_8 \Tr B_v[v\cdot A,[iD^\mu,A_\mu]]\bar B_v\nonumber\\
    && +h_1\Tr {\bar B}_v\chi_+ (iv\cdot{\cal D})B_v
       +h_2\Tr {\bar B }_v(iv\cdot{\cal D})B_v \chi_+
       +h_3\Tr {\bar B }_v(iv\cdot{\cal D})B_v \Tr \chi_+\nonumber\\
    && +l_1 \Tr \bar B_v [\chi_-, v\cdot A] B_v
       +l_2 \Tr \bar B_v B_v [\chi_-, v\cdot A]
       +l_3[\Tr \bar B_v \chi_-,\Tr B_vv\cdot A],
   \eeq
where the covariant derivative ${\cal D}_{\mu}$ for baryon fields is defined by
   \beq
      {\cal D}_{\mu}B_v &=& \partial_{\mu}B_v+[V_{\mu},B_v], \nonumber\\
      {\cal D}_{\mu}T_{v,abc}^{\nu} &=& \partial_{\mu}T_{v,abc}^{\nu}+
         	(V_{\mu})_a^dT^{\nu}_{v,dbc}+(V_{\mu})_b^dT_{v,adc}^{\nu}
                +(V_{\mu})_c^dT_{v,abd}^{\nu},
   \eeq
$\delta_T$ is the $SU(3)$ invariant decuplet-octet mass difference, and
   \beq
    \chi_{\pm} &\equiv& \xi {\cal M} \xi {\pm}
    \xi^\dagger {\cal M} \xi^\dagger,
   \eeq
with  ${\cal M}={\rm diag}\,(m_u,m_d,m_s)$
the quark mass matrix that breaks chiral symmetry explicitly. 
There are many other terms involving the decuplet that one can write down
but we have written only those that enter in the calculation. Among the many 
parameters that figure in the Lagrangian, a few can be fixed right away.
For instance, we will simply fix the constants $F$ and $D$ at tree order
since to ${\O} (Q^3)$ that we will be interested in, they are not modified.
We shall use $D=0.81$ and $F=0.44$. The constant $C$ can also be fixed at this
stage from the decay process  $\Delta(1230)\rightarrow N\pi$. We shall use 
$|C|^2(\approx 2.58)$. Of course, the flavor $SU(3)$ can be substantially broken
as we will discuss later, so one cannot take this value too seriously.
The determination of all other constants $a_i$,..., $l_i$ 
(or more precisely the combinations thereof) will be described below.

The number of parameters that seem to enter may appear daunting to some readers
but the situation turns out to be much simpler than what it looks.
As we will see later, once the constants are grouped into an appropriate
form, there remain only four parameters for on-shell $K^\pm N$ amplitudes.
These parameters can be fixed on-shell by the four s-wave scattering lengths.
Off-shell, however, one parameter remains free but the off-shell amplitude
turns out to be rather insensitive to the one free parameter.
This drastic simplification can be understood easily as follows.
First of all, the heavy-fermion formalism (in short HFF)
makes those subleading terms 
({\it i.e.}, terms with $\nu\geq 2$) involving
the spin operator $S_{\mu}$ vanish, since they are proportional to
$S\cdot q$, $S\cdot q'$, or $S\cdot q
S\cdot q'$, all of which are identically zero. As a consequence, there are
no contributions to the $s$-wave meson-nucleon scattering
amplitude from one-loop diagrams in which the external meson lines couple to
baryon lines through the axial vector currents. This leaves
only six topologically distinct one-loop diagrams, Fig.1, 
(out of thirteen in all) to calculate 
for the $s$-wave meson-nucleon scattering, apart from the usual
radiative corrections in external lines.
Since we are working to ${\cal O}(Q^3)$, only
${\cal L}^{(1)}$ enters into the loop calculation. Loops involving other
terms can contribute at $\O (Q^4)$ or higher. The next term ${\cal L}^{(2)}$
contributes terms at order $\nu=2$, that is, at tree order.
These will be determined by the $KN$ sigma term and terms that could be
calculated by resonance saturation. There are some uncertainties here as we 
shall point out later, but they turn out to be quite insignificant in the
results.  The next terms in ${\cal
L}^{(3)}$ remove the divergences in the one-loop
contributions and involve two finite counter terms -- made up of two linear
combinations of the many parameters appearing in the Lagrangian --
that are to be determined
empirically. As we will mention later, these constants are determined solely by
isospin-odd amplitudes, the loop contribution to isospin-even amplitudes being
free of divergences.


\section{KN Scattering Amplitudes}
\subsection{On-shell amplitudes}
\indent

The complete on-shell s-wave KN scattering amplitudes calculated to 
N$^2$LO ($\O (Q^3)$) \cite{LJMR} read
  \be
    a_0^{K^\pm p} &=& \frac{m_B}{4\pi f^2 (m_B+M_K)} 
    \left[
    \mp M_K
    + (\bar d_s+\bar d_v) M_K^2 +\left\{ (L_s+L_v) \pm (\bar g_s +\bar g_v)
    \right\} M_K^3 \right]
  \nonumber\\
&&+\delta a_{\Lambda^\star}^{K^\pm p}\nonumber\\
  a_0^{K^\pm n} &=& \frac{m_B}{4\pi f^2 (m_B+M_K)} 
    \left[ \mp \frac 12 M_K 
    + (\bar d_s-\bar d_v) M_K^2 +\left\{ (L_s-L_v) \pm (\bar g_s -\bar g_v) 
    \right\} M_K^3 \right]
  \label{scattamp}\ee
where 
$\bar d_s$ is the t-channel isoscalar contribution of ${\cal O}(Q^2)$, and
$\bar d_v$ is the t-channel isovector one of ${\cal O}(Q^2)$: 
  \be
    \bar d_s &=& -\frac{1}{2 B_0} (a_1+2 a_2 +4 a_3) + \frac 14 (d_1
    +d_2 +d_7 +d_8) +\frac 12 (d_3+d_4) +d_5 +d_6 
    \nonumber\\
    \bar d_v &=& -\frac{1}{2 B_0} a_1 + \frac 14 (d_1+d_2+d_7+d_8)
  \label{eqd}\ee
with $B_0=M_K^2/(\hat m+m_s)$ where $M_K$ is the kaon mass and $\hat{m}=
(m_u+m_d)/2$. Here $\delta a_{\Lambda^\star}^{K^\pm p}$ is the 
contribution from the $\Lambda^\star$ to be specified below and
$L_s$($L_v$) is the finite crossing-even t-channel isoscalar (isovector)
one-loop contribution
  \be
    L_s M_K &=& \frac{1}{128\pi f^2 M_K^2} \left( 
   \frac 13 (D-3F)^2 (M_\pi^2+3 M_\eta^2) M_\eta -9 M_K^2 \sqrt{M_\eta^2
   -M_K^2} \right)
    \nonumber\\
   &\approx& -0.109 \; \fm
    \nonumber\\
   L_v M_K &=& \frac{1}{128\pi f^2 M_K^2}\left( -\frac 13 (D+F) (D-3F) (M_\pi^2 +
   3 M_\eta^2) (M_\pi+M_\eta) -3M_K^2\sqrt{M_\eta^2-M_K^2} \right.
   \nonumber\\
   && \;\;\;\;\;\;\left. -\frac 16 (D+F)(D-3F) (M_\pi^2+3 M_\eta^2)
   (M_\pi^2+M_\eta^2) \int_0^1 \frac{1}{\sqrt{(1-x)M_\pi^2+x M_\eta^2}}
   \right)
   \nonumber\\
   &\approx& +0.021 \; \fm
   \ee
where $f=93 MeV$ and physical masses are used to obtain the numbers.
The quantity $\bar g_s (\bar g_v)$ is the crossing-odd t-channel isoscalar
(isovector) contribution from one-loop plus counter terms which after the
dimensional regularization specified in Appendix C, takes the form
  \be
  \bar g_{s,v} &=& \alpha_{s,v}^r +\beta_{s,v}^r 
    + \frac{1}{32\pi^2} \frac{1}{f^2 M_K^2}
   \left( \gamma_{s,v}+ \sum_{i=\pi,K,\eta} \delta_{s,v}^i
   \ln \frac{M_i^2}{\mu^2} \right)\label{gterms}
  \ee
where $\mu$ is the arbitrary scale parameter that enters in the dimensional
regularization and $\alpha$ and $\beta$ are contributions from 
the counter terms in ${\cal O}(Q^3)$,\footnote{On-shell, one can combine
$\alpha$ and $\beta$ into one set of parameter to be determined from
experiments.  Off-shell, however, they are multiplied by a different power
of the frequency $\omega$ as indicated in Appendix E and hence represent
two independent parameters. This introduces one unfixed parameter in the
off-shell case. However it turns out that the off-shell
amplitudes are rather insensitive to the precise values of these constants,
so we set (somewhat arbitrarily) $\alpha_{s,v}^r\approx \beta_{s,v}^r$ in
our calculation, Figure 2.}
and $\gamma$ and $\delta$ are finite loop contributions.
The explicit forms of $\alpha$, $\beta$, $\gamma$ and $\delta$
are given in Appendix C. It should be noted that while $\alpha$ and $\beta$ are
$\mu$-dependent, $\bar{g}$ is scale-independent. ($\gamma$ and $\delta$ are 
scale-independent numbers.) Thus if one fixes $\bar{g}$ from experiments, then
for a given $\mu$, one can fix $\alpha +\beta$ at a fixed $\mu$. 
Equivalently, we can separate out the specific $\mu$-dependent terms
so as to cancel
the $\ln \mu$ term in eq.(\ref{gterms}), thereby defining $\mu$-independent
constants $\alpha^\prime$ and $\beta^\prime$
\be
  \alpha_{s,v}^r +\beta_{s,v}^r =
   \alpha_{s,v}^\prime +\beta_{s,v}^\prime 
    - \frac{1}{32\pi^2} \frac{1}{f^2 M_K^2}
    \sum_{i=\pi,K,\eta} \delta_{s,v}^i
   \ln \frac{M_i^2}{\mu^2}\label{gpterms} 
\ee
and determine 
   $\alpha_{s,v}^\prime$ and  $\beta_{s,v}^\prime$ from experiments.
{}From now on when we go off-shell ({\ie}, in Appendix E), we will drop the 
primes understanding that we are dealing with the $\mu$-independent parameters. 
(On-shell, this subtlety is not relevant since we can work directly with
$\bar{g}$ of eq.(\ref{gterms}).)

To understand the role of the $\Lambda^\star$, we observe that the measured
scattering lengths are repulsive in all channels except 
$K^- n$ \cite{BS,lambdadata}:\footnote{
Although the experimental $K^- N$ scattering
lengths are given with error bars, the available $K^+ N$ data are not
very well determined. Since both are used in fitting the parameters of the
Lagrangian, we do not quote the error bars here and shall not
use them for fine-tuning. For our purpose, 
we do not need great precision in the data as the results are 
extremely robust against changes in the parameters.
}
  \be
  a_0^{K^+p} = -0.31 \fm,& \;\;\;\;& a_0^{K^-p} = -0.67 +i 0.63 \fm
  \nonumber\\
  a_0^{K^+n} = -0.20 \fm,& \;\;\;\;& a_0^{K^-n} = +0.37 +i 0.57 \fm .
\label{expscatt}
  \ee
The repulsion in $K^-p$ scattering cannot be explained from eq.(\ref{scattamp})
without the $\Lambda^\star$ contribution.
In fact it is well known that the contribution of the 
$\Lambda(1405)$ bound state
gives the repulsion required to fit empirical data for s-wave $K^-p$ 
scattering \cite{LJMR,lambda1405}. As mentioned, we may introduce the 
$\Lambda^\star$ as an elementary field. To the leading order in the 
chiral counting, it takes the form
  \beq
  {\cal L}_{\Lambda^\star}&=&
     \bar\Lambda^\star_v(iv\cdot\partial-m_{\Lambda^\star}+m_B)\Lambda^\star_v
     + \left( \sqrt{2} {g}_{\Lambda^\star}\:\Tr (\bar\Lambda^\star_v
     v\cdot A B_v) + {\rm h.\:c.}\right) .
  \eeq
The coupling constant $g_{\Lambda^\star}$ can be fixed by
the decay width $\Lambda^\star\rightarrow \Sigma\pi$ \cite{LJMR} if one ignores
$SU(3)$ breaking
\be
g^2_{\Lambda^\star} (pK^-)\approx g^2_{\Lambda^\star} 
(\Sigma\pi)\approx 0.15.\label{su3value}
\ee
This is what one would expect at tree order. If one wants to go
to one-loop order \cite{Savage} corresponding to $\O (Q^3)$ at which
$SU(3)$ breaking enters, then we encounter two counter terms $h^\star_{1,2}$,
  \be
 {\cal L}^{\nu=3} &=& 
  h^\star_1 \sqrt{2} \bar\Lambda^\star_v \Tr (\chi_+ v\cdot A B_v) 
   + h^\star_2 \sqrt{2} \bar\Lambda^\star_v \Tr (\chi_+ B_v v\cdot A) +h.c.
  \ee
and the renormalized coupling to ${\cal O}(Q^3)$ will take the form
   \be
   g^r_{\Lambda^\star} (\Sigma\pi) &=& g_{\Lambda^\star} 
   +  \sum_{i=\pi,K,\eta} \alpha_i^{\Sigma\pi} 
    \ln\frac{M_i^2}{\mu^2}
   +\beta^{\Sigma\pi} +2\; ({h^\star_1}^r+{h^\star_2}^r)\hat m
   \nonumber\\
   g^r_{\Lambda^\star} (pK^-) &=& g_{\Lambda^\star} 
   +\sum_{i=\pi,K,\eta} \alpha_i^{pK^-} 
    \ln\frac{M_i^2}{\mu^2}
   +\beta^{pK^-} +2\; ({h_1^\star}^r m_s+ {h_2^\star}^r \hat m)
   \label{renor2}\ee
where $\alpha_i$ and $\beta$ are calculable loop contributions. 
In \cite{Savage},
Savage notes that if one ignores the counter terms, then the finite log terms
would imply that $g_{\Lambda^\star} (pK^-)$ would 
come out to be considerably smaller
than $g_{\Lambda^\star} (\Sigma\pi)$, presumably due to an $SU(3)$ breaking.
In our approach we choose to pick the constants from experiments since we 
see no reason to suppose that the counter terms are zero and furthermore
the presently available data\cite{lambdadata} give
\be
(g^2_{\Lambda^\star} (pK^-))_{exp}\approx 0.25\label{expcoup}
\ee
which is {\it bigger} than the $SU(3)$ value (\ref{su3value}). We will take this
value in our calculation. It should also be mentioned that the Callan-Klebanov
skyrmion predicts a value close to (\ref{su3value}) \cite{Scoccola}.
The $\Lambda^\star$ contribution to the kaon-proton scattering amplitude 
is now completely determined,
   \be
   \delta a_{\Lambda^\star}^{K^\pm p} &=& - \frac{m_B}{4\pi f^2 (m_B+M_K)} 
   \left[ \frac{ g_{\Lambda^\star}^2 M_K^2}{m_B \mp M_K -
m_{\Lambda^\star}}\right].
   \label{Lama}\ee
To one-loop order, the $\Lambda^\star$ mass picks up an imaginary part
through the graph $\Lambda^\star\rightarrow \Sigma\pi\rightarrow \Lambda^\star$.
In our numerical work we will take $m_{\Lambda^\star}$ to be complex.
The presence of the imaginary part explains that the empirical coupling 
constant (\ref{expcoup}) is bigger than the $SU(3)$ value (\ref{su3value}).

We are left with four parameters in (\ref{scattamp}), 
$\bar d_s,\bar d_v, \bar g_s$ and $\bar g_v$, which we can determine
with the four experimental (real parts of) scattering lengths (\ref{expscatt}). 
The results are
   \be
   \bar d_s \approx 0.201  \fm, \;\; && \bar d_v \approx 0.013 \fm,
   \nonumber\\
   \bar g_s M_K \approx 0.008 \fm, \;\; && \bar g_v M_K \approx 0.002 \fm.
   \label{num}\ee
The scattering amplitudes in each chiral order are
given in Table 1.
One sees that while the order $Q$ and order $Q^2$ terms are comparable,
the contribution of order $Q^3$ is fairly suppressed relative to them.
As a whole, the subleading chiral corrections are verified to be
consistent with the ``naturalness" condition as required
of effective field theories. 
Using other sets of values of $f$, $D$ and $F$ does not change
$L_s$ and $L_v$ significantly and leave unaffected  our main conclusion.
As expected, the $\Lambda^\star$ plays a predominant role in $K^- p$
scattering near threshold. This indicates that it will be essential
in describing kaon-nuclear interactions, e.g., kaonic atoms.

\subsection{Off-Shell Amplitudes}
\indent

We now turn to off-shell s-wave $K^-$ forward scattering off static nucleons.
The kinematics involved are $t=0$, $q^2=q'^2=\omega^2$, $s=(m_B+\omega)^2$ with
an arbitrary (off-shell) $\omega$. (See Appendix E.)

In going off-shell, we need to separate different kinematic dependences
of the constant $\bar d_{s,v}$ of eq.(\ref{eqd}) which consists of what would correspond
to the KN sigma term $\Sigma_{KN}$ at tree order, $\sigma_{KN} (\approx
\Sigma_{KN}) \equiv -\frac 12 (\hat m +m_s) (a_1 +2 a_2 +4a_3)$ involving the
quark mass matrix $\M$, and the $ d_i$ terms containing two time derivatives.
The $\sigma_{KN}$ term, which gives an attraction, is a constant 
independent of the kaon frequency $\omega$,
whereas the $d_i$ terms which are repulsive
are proportional to $\omega^2$ in s-wave. In kaonic atoms and kaon 
condensation, the $\omega$ value runs down from
its on-shell value $M_K$. This means that as $\omega$ goes down, the attraction
stays unchanged and the repulsion gets suppressed.
Thus while for on-shell amplitudes, they can be obtained independently
of any assumptions, they need to be separated for off-shell amplitudes
that we are interested in. If we could determine the KN sigma term from
experiments, there would be no ambiguity. The trouble is that the sigma
term extracted from experiments is not precise enough to be useful.
The presently available value ranges 
\be
\Sigma_{KN}\sim (200 - 400)\ {\rm MeV}.
\ee
Here we choose to separate the two components by estimating the contributions
to $ d_i$ from the leading $1/m_B$ corrections with the octet and
decuplet intermediate states in the relativistic Born graphs. As suggested
by the authors of ref.\cite{meissner} for $\pi N$ scattering, 
we might assume that the counter terms $ d_i$ could equally be
saturated by such intermediate states. This strategy is discussed in detail
in Appendix D. This is somewhat like saturating
the dimension-four counter terms $L_i$ in the chiral Lagrangian by
resonances, a prescription which turns out to be surprisingly successful. 
The reason for
believing that this might be justified in the present case is that
the $\O (Q^2)$ terms are not affected by chiral loops, so must represent
the degrees of freedom that are integrated out from the effective 
Lagrangian. But there are no known mechanisms that would contribute to
$ d_i$ other than the baryon resonances.
When computed by resonance
saturation, the contributions go like $1/m_B$. 
However this is not to be taken as $1/m_B$ corrections that
arise as relativistic corrections to the static limit of a relativistic
theory. The HFF as used in chiral perturbation theory does not
correspond merely to a non-relativistic reduction although at low orders,
they are equivalent. To be more specific, imagine starting with the
following relativistic Lagrangian density
   \be
   {\cal L}_{rel} &=& \cdots +  e_i \Tr[\bar B \gamma_\mu A^\mu 
   \gamma_\nu A^\nu B] +\cdots
    +    f_i \Tr[ D_\mu\bar B A^\mu A^\mu D_\mu B]+\cdots .
   \label{lagD}\ee
In going to the heavy-baryon limit, we get the
${\cal O}(Q^2)$ terms of the form
   \be
   {\cal L}_v &=& \cdots + d_i \Tr[\bar B_v v\cdot A^2 B_v] +\cdots
   \ee
with $ d_i=(d_{\frac{1}{m}}+ e_i +m_B^2 f_i +\cdots )$ where
$d_{\frac{1}{m}}$ is the calculable $1/m_B$ correction from the relativistic
leading-order Lagrangian. Clearly the $e_i$ and $f_i$ terms cannot be
computed \cite{Savage}. Thus if one imagines that the constants $d_i$ are
infested with the terms of the latter form, there is no way that one can
estimate these constants. While this introduces an element of uncertainty in our
calculation, it does not seriously diminish the predictivity of the theory:
Much of the uncertainty are eliminated in our determination of the parameters
by experimental data.
From eq.(\ref{num}) and the $ d_i$ terms estimated in Appendix D, we can 
extract the parameter\footnote{This is not the {\it sigma term}
$\Sigma_{KN}=\frac 12 (\bar{m}+m_s)\langle P|\bar{u}u+\bar{s}s|P\rangle$.
There are loop corrections to be added to this value.} 
\be
 \sigma_{KN} \approx 2.83 M_\pi.
\ee
This {\it parameter} will be used for off-shell KN scattering amplitudes 
and kaon self-energy.

The predicted off-shell $K^- p$ and $K^-n$ scattering amplitudes
are shown in solid line in Figure 2
for the range of $\sqrt{s}$ from $1.3$ GeV to $1.5$ GeV with
${g}_{\Lambda^\star}^2=0.25$ and $\Gamma_{{\Lambda^\star}}=50$ MeV taken from
experiments. (The dotted lines in Fig.2 are explained in Appendix D.)
The explicit formulas are listed in Appendix E. The 
$K^- n$ scattering is independent of the $\Lambda^\star$ and so the amplitude
varies smoothly over the range involved.\cite{LJMR}
Our predicted $K^-p$ amplitude is found to be in fairly good agreement with
the empirical fit of ref.\cite{Steiner}.
The striking feature of the real part of the $K^-p$ amplitude,
repulsive above and attractive
below $m_{\Lambda^\star}(1405 MeV)$ as observed here, 
and the $\omega$-independent
attraction of the $K^- n$ amplitude are relevant to kaonic atoms
\cite{kaonicatom} 
and to kaon condensation in ``nuclear star" matter. The imaginary 
part of the $K^- p$ amplitude is somewhat too high compared with the empirical
fits. This may have to do with putting the experimental
$\Lambda^\star$ decay width for the imaginary part of the mass. 
Self-consistency between loop corrections and the imaginary part of
the mass would have to be implemented to get the correct imaginary
part of the $K^- p$ amplitude.

\subsection{Adler soft-meson conditions}
\indent

The off-shell amplitude
calculated here does not satisfy Adler's soft-meson conditions that follow from
the usual PCAC assumption that the pseudoscalar meson field $\pi$
interpolate as the
divergence of the axial current. The chiral Lagrangian used here does not
give the direct relation $\pi^i \sim \del_\mu J_5^{i\mu}$ where $J_5^{i\mu}$
is the axial current with flavor index $i$. Therefore in the soft-meson limit
which corresponds in the present case to setting $\omega$ equal to zero,
the $\pi N$ amplitude is not given by $-\frac{\Sigma_{\pi N}}{f^2}$ as it
does in the case of Adler's interpolating field \cite{adler}. In fact,
it gives $\frac{\Sigma_{\pi N}}{f^2}$ which has the opposite sign to
Adler's limit. This led several authors to raise the possibility that
a different physics might be involved in the chiral perturbation description
of the off-shell processes that take place near $\omega=0$ \cite{YNMK,lutz}.

A simple answer  to this issue is that physics should not depend upon the
interpolating field for the Goldstone bosons $\pi$ \cite{CCWZ}. 
The physics is equivalent
whether one uses the $\pi$ field as defined by the chiral Lagrangian used
here or the $\pi^\prime\propto \del_\mu J_5^\mu$ field that gives Adler's 
conditions. Both
are interpolating fields and they are just field-redefinitions of each other.
This is natural since the Goldstone boson field is an auxiliary field in
QCD. An explicit illustration of the equivalence in the case considered in
this paper is given
in Appendix F, to which skeptics are referred. Let it suffice here to say
that if one wishes, one could rewrite the effective Lagrangian in such a way
that Ward-Takahashi identities, to which Adler's conditions belong, are
satisfied, without changing the physics involved. See ref. \cite{GSS}.


\section{Kaon Self-Energy and Kaonic Atom}
\indent

If one were to limit oneself to linear density approximation which 
would be reliable in dilute systems, then what we have obtained so
far is sufficient for studying kaon-nucleon interactions in many-body systems.
The off-shell KN amplitude calculated to $\O (Q^3)$ in impulse approximation
gives the optical potential, which, expressed as a self-energy,
is depicted by Fig.3a,
    \be
    \Pi^{imp}_K(\omega) &=& -\left( \rho_p  {\cal T}^{K^-p}_{free}(\omega)
        +\rho_n  {\cal T}^{K^-n}_{free}(\omega) \right)\label{self1}
    \ee
where ${\cal T}^{KN}$ is the off-shell s-wave KN transition matrix
\footnote{The amplitude ${\cal T}^{KN}$ taken on-shell, {\it i.e.},
$\omega=M_K$, and the scattering length
$a^{KN}$ are related by $ a^{KN} = \frac{1}{4\pi (1+M_K/m_B)}
{\cal T}^{KN} $.}. 
For the purpose of studying kaon condensation, the linear density approximation
may not be reliable enough and one would have to study the 
{\it effective action}
(or {\it effective potential} in translationally invariant systems).

To go beyond the linear density approximation, there are two major effects
to be considered. The first is the Pauli correction and the second is
many-body correlations.

The Pauli effect can be most straightforwardly taken into account
in the self-energy
by modifying the nucleon propagator in the loop graphs contributing
to the KN scattering amplitude to one appropriate in medium
    \be
    G^0(k) \simeq \frac{i}{v\cdot k +i\epsilon}
    -2\pi\delta(k_0)\theta (k_F -|\vec k|)
    \label{prop}\ee
where $k_F$ is the nucleon Fermi momentum related to density $\rho_N$ by
the usual relation
    $\rho_N =\frac{\gamma}{6\pi^2} k_{F_N}^3$
with the degeneracy factor $\gamma=2$ for neutron and proton in nuclear matter.
The resulting correction denoted
$\delta {\cal T}_{\rho N}^{K^- N}$ and given explicitly in
Appendix G is clearly nonlinear in density
and repulsive as befits a Pauli exclusion effect.

For the second effect, the most important one is the correlation involving
``particle-hole" excitations. This is of typically many-body nature.
There are two classes of correlations one would have to consider.
One involves non-strange particle-hole excitations and the other
strange particle-nonstrange hole excitations. All these can be mediated
by four-Fermi interactions described above.

We first consider the latter. These are depicted in Fig. 4.
Since we are dealing with s-wave kaon interaction, the most important
configuration that $K^-$ can couple to is the $\Lambda^\star$ particle-nucleon
hole (denoted as ${\Lambda^\star}  N^{-1}$ with $N$ either a proton ($p$) or neutron
($n$)). 
We shall return to nonstrange particle-hole correlations 
when we treat density effects on the basic constants of the Lagrangian
({\ie}, ``BR scaling").
Here we focus on the former type. Now for the s-wave in-medium kaon 
self-energy, the relevant
four-Fermi interactions that involve a $\Lambda^\star$ can be reduced to a
simple form involving two unknown constants
    \be
    {\cal L}_{4-fermion} &=& 
   C_{\Lambda^\star}^S \bar{\Lambda}^\star_v \Lambda^\star_v 
  \Tr \bar B_v B_v +  C_{\Lambda^\star}^T 
  \bar{\Lambda}^\star_v \sigma^k \Lambda^\star_v
     \Tr \bar B_v \sigma^k B_v
    \ee
where $C_{\Lambda^\star}^{S,T}$ are the 
dimension $-2$ ($M^{-2}$) parameters to be fixed
empirically and $\sigma^k$ acts on baryon spinor.

Additional (in-medium) two-loop graphs that involve ${\Lambda^\star} N^{-1}$
excitations are given in Fig. 4c. They do not however involve
contact four-Fermi interactions, so are calculable unambiguously.

We shall denote the sum of these contributions from Figs. 4
to the self-energy by $\Pi _{\Lambda^\star}$. A simple calculation gives
    \be
    \Pi_{\Lambda^\star}(\omega) &=& - \frac{g_{\Lambda^\star}^2}{f^2}
        \left(\frac{\omega}{\omega+m_B-m_{\Lambda^\star}} \right)^2
    \left\{ C_{\Lambda^\star}^S \rho_p \left( \rho_n +\frac 12 \rho_p \right)
    -\frac 32 C_{\Lambda^\star}^T \rho_p^2 \right\}
    \nonumber\\
    && -\frac{g_{\Lambda^\star}^4}{f^4} \rho_p
    \left(\frac{\omega}{\omega+m_B-m_{\Lambda^\star}} \right)^2
    \omega^2 \left( \Sigma_K^p (\omega) +\Sigma_K^n (\omega) \right)
    \label{pilambda}\ee
where ${g}_{\Lambda^\star}$ is the renormalized $KN\Lambda^\star$ coupling
constant determined in \cite{LJMR} and $\Sigma_K^N (\omega)$ is given by
    \be
    \Sigma_K^N (\omega) &=& \frac{1}{2\pi^2}\int_0^{k_{F_N}} 
    d|\vec k| \frac{|\vec k|^2}{ \omega^2-M_K^2-|\vec k|^2}.
    \ee
In eq.(\ref{pilambda}),
the first term comes from the diagrams of Figs. 4a and 4b
and the second term  from the diagram of Fig. 4c.
While the second term gives repulsion corresponding to a Pauli quenching,
the first term can give either attraction or repulsion depending on
the sign of $(C_{\Lambda^\star}^S [\rho_n+\frac 12 \rho_p]-
\frac 32 C_{\Lambda^\star}^T\rho_p)$ with the constants $C_{\Lambda^\star}^{S,T}$
being the only parameters that are not determined by
on-shell data.
 
The complete self-energy to in-medium two-loop order
is then
    \be
    \Pi_K(\omega) &=& -\left( \rho_p  {\cal T}^{K^-p}_{free}(\omega)
        +\rho_n  {\cal T}^{K^-n}_{free}(\omega) \right)
        - \left(\rho_p  \delta {\cal T}^{K^-p}_{\rho_N}(\omega)
        +\rho_n \delta {\cal T}^{K^-n}_{\rho_N}(\omega) \right)
        +\Pi_{\Lambda^\star}(\omega)\label{self2}.
    \ee

The additional parameters $C_{\Lambda^\star}^{S,T}$
that are introduced at the level of four-Fermi
interactions in the strange particle-hole sector require experimental
data involving nuclei and 
nuclear matter. We shall now discuss how these constants can
be fixed from kaonic atom data. In order to fix both of these constants,
we would need data over a wide range of nuclei. One sees in (\ref{pilambda})
that for the symmetric matter, what matters is the combination
$(C^S_{\Lambda^\star}-C^T_{\Lambda^\star})$. At present, this is the only combination
that we can hope to pin down from kaonic atom data. That leaves one 
parameter unfixed. We shall pick $C^S_{\Lambda^\star}$ for the
reason to be explained later. We shall parametrize
the proton and neutron densities by the proton fraction $x$
and the nucleon density $u=\rho/\rho_0$ as
    \be
    \rho_p = x\rho\; ,\;\; \rho_n =(1-x) \rho\; ,\;\; \rho = u\rho_0.
    \ee
Now what we know from
the presently available kaonic atom data \cite{kaonicatom} is that the optical
potential for the $K^-$ in medium has an attraction of the order of
    \be
    \Delta V \approx -(180\pm 20)\ \ {\rm  MeV}\ \ at \;\; u=0.97.
    \ee
This implies approximately for $x=1/2$
    \be
     ( C_{\Lambda^\star}^S -C_{\Lambda^\star}^T) f^2\approx 10.
\label{cvalue}
    \ee
Table 2 gives details of how this value is arrived at. It also lists the 
contributions of each chiral order to the self-energy (\ref{self2}), 
 $\Delta V=M_K^\star-M_K$ and $M_K^\star$ which we shall loosely call
``effective kaon mass" \footnote{This is not, strictly speaking, a mass
but we shall refer to it as such in labeling the figures.}
\be
M_K^\star \equiv\sqrt{M_K^2+\Pi_K}. 
\ee
To exhibit the role of $\Lambda^\star$ in the kaon self-energy, we list
each contribution of $\Pi$. Here $\Pi_{free}=
-\rho_N {\cal T}^{K^-N}_{free}$, $\delta\Pi = -\rho_N
\delta {\cal T}^{K^-N}$, 
$\Pi_{\Lambda^\star}^1$  corresponds to the first term
of eq.(\ref{pilambda}) which depends on $C_{\Lambda^\star}^{S,T}$  
and $\Pi_{\Lambda^\star}^2$
to the second term independent of $C_{\Lambda^\star}^{S,T}$.
We observe that the $C_{\Lambda^\star}^{S,T}$-dependent term plays a crucial 
role for attraction in kaonic atom. 

In Table 3 and Fig. 5, we list the predicted density dependence of 
the real part of
the kaonic atom potential for $x=0.5$ obtained for
$(C_{\Lambda^\star}^S -C_{\Lambda^\star}^T) f^2 \approx 10$.

To understand what the remaining parameter $C^S_{\Lambda^\star}$ is physically,
we consider the mass shift of the $\Lambda^\star$ in medium. To one-loop
order, there are two graphs given in Fig.6. A simple calculation gives
\be
\delta m_{\Lambda^\star}=\sum_{i=a,b}\delta 
 \Sigma^{(i)}_{\Lambda^\star} (\omega=m_{\Lambda^\star}-m_B)
\ee
where
\be
\delta\Sigma_{\Lambda^\star}^{(a)} (\omega) 
&=&-\frac{g_{\Lambda^\star}^2}{f^2}\omega^2\left(
\Sigma^p_K (\omega) +\Sigma^n_K (\omega)\right)\nonumber\\
\delta\Sigma_{\Lambda^\star}^{(b)} (\omega) &=& 
-C_{\Lambda^\star}^S (\rho_p+\rho_n).
\ee
The superscript $(a,b)$ stands for the figures (a) and (b) of Fig. 6.
The contribution from Fig.6a is completely given with the known constants.
The dependence on the unknown constant $C^S_{\Lambda^\star}$ appears linearly in the
Fig.6b. For the given values adopted here, the mass shift is numerically
\be
\delta m_{\Lambda^\star} (u,x,y)=[r(u,x)-150.3\times u\times y] \ \ {\rm MeV}
\label{lambdamass}
\ee
where  $y=C^S_{\Lambda^\star} f^2$ and $r(u,x)\equiv\delta \Sigma_{\Lambda^\star}^{(a)}$ 
with the numerical values given in Table 4.

One can see from eq.(\ref{lambdamass}) and Table 4 that the shift in the 
$\Lambda^\star$ mass in medium is primarily controlled by the constant
$C^S_{\Lambda^\star}$. For symmetric matter ($x=1/2$, $u=1$), the shift is 
zero for $y=0.41$ and linearly dependent on the $y$ value for non-symmetric
matter. At present
we have no information as to whether the medium lowers or raises the mass
of the $\Lambda^\star$, so it is really a free parameter but it is
reasonable to expect that if any the shift cannot be very significant.
We consider therefore that  a reasonable value for $y$ is $\O (1)$.

In Table 5 and Fig.5 are given the properties of $K^+$ in nuclear matter.
The self-energy of $K^+$ is simply obtained from that of $K^-$ by crossing
$\omega\rightarrow -\omega$.
One can note here that the interaction of $K^+$ with nuclear medium is
quite weak as predicted by phenomenological models and supported by 
experiments. The $M_K^\star$ grows slowly as a function of density.\footnote{
Recent heavy-ion experiments at Brookhaven\cite{stachel} find that $K^+$'s 
come out of quark-gluon plasma with a temperature of order of 20 MeV which
is much lower than the freeze-out temperature of $\sim 140$ MeV. To understand
this phenomenon, there has to be a mechanism at high temperature and density
that reduces the $K^+$ mass considerably. Within the framework adopted here,
this can happen only if the strong repulsion due to the $\omega$ exchange
lodged in the first term of ${\cal L}^{(1)}$, eq.(\ref{l1}), is strongly 
suppressed so that the attraction coming from the sigma term becomes operative,
thereby reducing the mass. As discussed in a recent paper by Brown and Rho
\cite{br94}, this can indeed happen if the vector meson decouples at high 
temperature and density. Here we are not concerned with this regime.} 
This is a check of the consistency of the chiral expansion approach to 
kaon-nuclear interactions.


\section{Critical Density for Kaon Condensation}
\indent

We have now all the ingredients needed to calculate the critical density
for negatively charged kaon condensation in dense nuclear star matter.
For this, we will follow the procedure given in
\cite{TPL}. As argued in \cite{BKR}, we need not consider pions
when electrons with high chemical potential can trigger condensation through
the process $e^-\rightarrow K^- \nu_e$. Thus we can focus on the spatially
uniform condensate
    \be
    \langle K^-\rangle =v_K e^{-i\mu t}
    \ee
where $\mu$ is the chemical potential which is equal, by Baym's theorem
\cite{baym}, to the electron chemical potential. 
The energy density $\tilde\epsilon$ -- which is related to the
effective potential in the standard way -- is given by,
    \be
    \tilde \epsilon (u,x,\mu, v_K) &=& \frac 35 E_F^{(0)} u^{\frac 53} \rho_0
        +V(u) +u\rho_0 (1-2x)^2 S(u) \nonumber\\
    &&-[\mu^2 -M_K^2 -\Pi_K (\mu,u,x)]
        v_K^2+ \sum_{n\ge 2} a_n(\mu,u,x) v_K^n \nonumber\\
    && +\mu u\rho_0 x +\tilde\epsilon_e +\theta(|\mu|-m_\mu)\tilde \epsilon_\mu
    \label{effen}\ee
where $E_F^{(0)}=\left( p_F^{(0)}\right)^2/2m_B$ and 
$p_F^{(0)}=(3\pi^2\rho_0 /2)^{\frac 13}$ are, respectively,
 Fermi energy and momentum at nuclear density. The
$V(u)$ is a potential for symmetric nuclear matter
as described in \cite{PAL} which is presumably subsumed in contact four-Fermi
interactions (and one-pion-exchange -- nonlocal -- interaction)
in the non-strange sector as mentioned above. It will affect
the equation of
state in the condensed phase but not the critical density, so we will
drop it from now on. The nuclear symmetry energy $S(u)$ -- also
subsumed in four-Fermi interactions in the non-strange sector -- does
play a role as we know from \cite{PAL}: Protons enter to neutralize the
charge of condensing $K^-$'s making the resulting compact star
``nuclear" rather than neutron star as one learns in standard astrophysics
textbooks. We take the form advocated in \cite{PAL}
    \be
    S(u) &=& \left(2^{\frac 23}-1\right) \frac 35 E_F^{(0)}
        \left(u^{\frac 23} -F(u) \right) +S_0 F(u)
    \ee
where $F(u)$ is the potential contributions to the symmetry energy and
$S_0 \simeq 30 MeV$ is the bulk symmetry energy parameter.
We use three different forms of $F(u)$ as in \cite{PAL}
    \be
    F(u)=u\;,\;\; F(u) =\frac{2u^2}{1+u}\;,\;\; F(u)=\sqrt u.
    \label{SE}
    \ee
It will turn out that
the choice of $F(u)$ does not significantly affect the critical density.
The contributions of the filled Fermi seas of electrons and muons 
are\footnote{We ignore hyperon Fermi seas in this calculation. We do not
expect them to be important for s-wave kaon condensation.}
\cite{TPL}
    \be
    \tilde \epsilon_e &=& -\frac{\mu^4}{12\pi^2} \nonumber\\
    \tilde \epsilon_\mu &=& \epsilon_\mu -\mu \rho_\mu 
    = \frac{m_\mu^4}{8\pi^2}\left((2t^2+1) t\sqrt{t^2+1}
    -\ln(t^2+\sqrt{t^2+1}
    \right) -\mu \frac{p_{F_\mu}^3}{3\pi^2}
    \ee
where $p_{F_\mu} =\sqrt{\mu^2-m_\mu^2}$ is the Fermi momentum and $t=p_{F_\mu}
/m_\mu$.

The ground-state energy prior to kaon condensation
is obtained by extremizing the energy density $\tilde\epsilon$
with respect to  $x$, $\mu$ and $v_K$:
    \be
    \left. \frac{\partial\epsilon}{\partial x}\right |_{v_K=0}=0 \;,\;\;
    \left. \frac{\partial\epsilon}{\partial \mu}\right |_{v_K=0}=0 \;,\;\;
    \left. \frac{\partial\epsilon}{\partial v_K^2}\right |_{v_K=0}=0
    \ee
from which we obtain three equations corresponding, respectively,
to beta equilibrium, charge neutrality and dispersion relation:
    \be
    \mu &=& 4 (1-2 x) S(u) \nonumber\\
    0 &=& -x u\rho_0 +\frac{\mu^3}{3\pi^2}
         +\theta(\mu-m_\mu)\frac{p_{F_\mu}^3}{3\pi^2} \nonumber\\
    0 &=& D^{-1} (\mu, u,x) =\mu^2-M_K^2 -\Pi_K(\mu,u,x)\equiv \mu^2-
    {M^\star_K}^2 (\mu,u,x).
    \ee
The proton fractions $x(u)$ and chemical potentials $\mu$ prior 
to kaon
condensation are plotted in Fig. 7 and Figs. 8 $\sim$ 10 for various choices of
the symmetry energy $F(u)$.
We have solved these equations using
for the kaon self-energy (a) the linear density approximation, eq.(\ref{self1})
and (b) the full two-loop result, eq.(\ref{self2}). Table 6.(a) shows the case (a)
for different symmetry energies eq.(\ref{SE}). We see that the precise form of
the symmetry energy does not matter quantitatively.
The corresponding ``effective kaon mass" ${M^\star_K}$
is plotted vs. $u$ in Figs. 8 $\sim$ 10  in solid line. Note that even in this linear
density approximation kaon condensation {\it does} take place, {\it albeit} at
a bit higher density than obtained before.

For the value that seems to be required by the kaonic atom
data, (\ref{cvalue}),
the critical density comes out to be about $u_c\approx 3$, rather close
to the original Kaplan-Nelson value.

In Table 6.(b) and Figs. 8 $\sim$ 10 are given 
the predictions for a wide range of
values for $C_{\Lambda^\star}^S f^2$. What is remarkable here is
that while the $C_{\Lambda^\star}^{S,T}$-dependent four-Fermi interactions are 
{\it essential} for triggering kaon condensation,
the critical density is quite insensitive to their strengths.
In fact, as one can see in Table 7, reducing the constant $(C_{\Lambda^\star}^S-C_{\Lambda^\star}^T)f^2$
that represents the kaonic atom attraction
by an order of magnitude to 1 with 
$C_{\Lambda^\star}^S f^2=10,\ 0$ modifies the
critical density only to $u_c\approx 3.3,\ 4.5$, respectively.

To see how robust kaon condensation is with respect to 
the $\Lambda^\star KN$ coupling constant, let us take the  
extreme value $g_{\Lambda^\star}^2 \approx 0.05$ used in
\cite{Savage} which gives the 
wrong sign to the $K^- p$ amplitude at threshold. 
In Table 8, $\Delta V$ of kaonic atom ($x=0.5$) is given for $u=0.97$ and 
for various choices of $(C^S_{\Lambda^\star}-C_{\Lambda^\star}^T) f^2$.
Constraining to the kaonic atom data implies approximately -- within the 
range of error involved --
$ (C_{\Lambda^\star}^S-C_{\Lambda^\star}^T) f^2 \approx 70 $.
In Table 9, the resulting critical densities are given for
$(C_{\Lambda^\star}^S-C_{\Lambda^\star}^T) f^2 \approx 70$, and those 
for other sets of $C_{\Lambda^\star}^S$ and $C_{\Lambda^\star}^T$ in Table 10.
We see that given the constraint from kaonic atom data,
the resulting critical densities are sensitive neither to 
the value of $C_{\Lambda^\star}^{S,T}$ nor to  $g_{\Lambda^\star}$.
   

\section{Four-Fermion Interactions and ``Scaled" Chiral Lagrangians}
\indent

So far we have ignored four-Fermi interactions in the non-strange sector
with the understanding that the effects not involving strangeness are to be
taken from what we know from nuclear phenomenology. From the chiral Lagrangian
point of view, this is not satisfactory. One would like to be able to
describe both the non-strange and strange sectors {\it on the same footing}
starting from a three-flavor chiral Lagrangian. While recent  developments
indicate that nuclear forces may be understood at low energies in terms of
a chiral Lagrangian \cite{wein,vankolck}, it has not yet been possible to
describe the ground-state property of nuclei including nuclear matter
starting from a Lagrangian that has explicit chiral symmetry. So the 
natural question is: What about many-body correlations in the 
non-strange sector, not to mention those in the strange sector for which
we are in total ignorance?

Montano, Politzer and Wise \cite{MPW} addressed a related question in
pion condensation in the chiral limit and found that four-Fermi
interactions in the non-strange sector
played an important role in inhibiting p-wave pion
condensation in dense matter. It has of course been known since some time
that the mechanism that quenches $g_A$ in nuclear matter to a value
close to unity banishes the pion condensation density beyond
the relevant regime. As shown in \cite{elaf}, this mechanism can be 
incorporated by means of a four-Fermi
interaction in a chiral Lagrangian in a channel corresponding to
the spin-isospin mode (that is, the Landau-Migdal $g^\prime$ interaction).

In this section, we discuss a simple approach to including the main correlations
in kaon condensation. We cannot do so in full generality to all orders
in density but we can select what we consider to be the dominant ones by
resorting to the scaling argument (which we shall
call ``BR scaling") introduced by Brown and Rho \cite{br94,br91}. 

How to implement the BR scaling in higher-order chiral expansion with the 
multiple scales that we are dealing with
has not yet been worked out. It has up to date been formulated so as to be
implemented {\it only at tree order} with the assumption 
that once the BR scaling is incorporated, higher-order terms are naturally
suppressed. If this assumption is valid, which we can check \`a posteriori,
by taking the BR scaling
into account at tree order, that is, at $\O (Q^2)$ in our 
case, we will be including most of higher-order density dependences
through the simple scaling. 

We first consider the leading-order ($\O (Q)$) term, which
for s-wave kaon-nucleon interactions is given by the first term of
eq.(\ref{l1}) 
\be
\sim \frac{1}{f^2} K^\dagger\del_0 K B_v^\dagger B_v.\label{weinbergterm}
\ee
In terms of vector-meson exchanges, it is equivalent to an 
$\omega$-meson exchange,
attractive in the $K^-$ channel and repulsive in the $K^+$ channel.
Four-Fermi interactions in the non-strange channel can modify this interaction,
the most important one being
\be
\sim D \frac{1}{f^2} K^\dagger\del_0 K (B_v^\dagger B_v)^2
\ee
with $D$ an unknown constant. One may try to estimate the constant $D$ using
dynamical models. For instance, one can have a four-Fermi interaction that
arises when a massive scalar ($\sigma$ in the linear $\sigma$ model)
is integrated out. This would increase the attraction of 
eq.(\ref{weinbergterm})
in the $K^-$ channel, which is equivalent to increasing
the magnitude of $D$. One can also
have a four-Fermi interaction that arises from 
integrating out a massive
vector exchange of the $\omega$-meson quantum number, 
responsible for
the screening of the attraction by a factor $(1+F_0)^{-1}$ where $F_0$ is
the Landau-Migdal parameter $>0$ in the density regime we are
interested in \cite{pethick}. This would decrease the magnitude of $D$. 
There are of course other terms but 
if we add them all up
and write an effective two-Fermi interaction with correct symmetries
by taking the mean-field
$\langle \bar{B}_v B_v\rangle\approx \rho$, then the higher-density dependence
is expected to modify (\ref{weinbergterm}) to
\be
\sim \frac{1}{{f^\star}^2} K^\dagger\del_0 K B_v^\dagger B_v \label{weinterm}
\ee
where the asterisk denotes density dependence $f^\star=f(\rho)$. Later we will 
assume this to be given by the BR scaling \cite{br94,br91}. Whether or not
this assumption is viable will be tested {\it \`a posteriori}.

Similar four-Fermi interaction
terms can also be written down for the $\O (Q^2)$ terms. 
However the effect here is expected to be less important for the
following reason. First of all,
the term involving the KN sigma term is associated with
the strange quark property, in particular with the kaon decay constant
$f_K$  and we expect that the $f_K$ is not modified significantly 
since the s-quark condensate does not change much
as density and/or temperature is increased\cite{Kogut}. 
This implies that 
four-Fermi interactions will be less effective in modifying this term.
The same must be the case with the counter terms $\bar d_i$ that are 
associated with
the octet and decuplet of strange-quark flavor. We may assume them to
be also unaffected by the renormalization. 

In sum, {\it our proposition is that 
the net effect of multi-Fermi interactions in the non-strange sector
can be summarized by
the scaling $f\rightarrow f^\star$ in the leading term in (\ref{l1}).}
As stated, we propose this scaling to be given by  the BR scaling
\be
\frac{f^\star}{f}\approx \frac{m_V^\star}{m_V}\approx 1-cu
\ee
with $c\approx 0.15$. Since the multi-Fermi interactions are higher order
in the chiral counting in terms of an expansion $k_F/\Lambda_\chi$, it is 
consistent to leave them out in the loop corrections, that is at $\O (Q^3)$.

The self-energy for kaonic atoms with the BR scaling is given in dashed line
in Fig. 5 and tabulated in Tables 11 and 12.
Comparing with Tables 2 and 3, we see that the BR scaling gives 
only a slightly more attractive potential at low densities.
The attraction, however, increases significantly at higher densities.
{}From Table 11, we find for kaonic atoms, approximately,
\be
(C_{\Lambda^\star}^S-C_{\Lambda^\star}^T) f^2 \approx 10.
\ee
This is roughly the same as without the BR scaling since the scaling effect
is not important at $u=0.97$. The BR scaling becomes important in the region
where condensation sets in. The corresponding
critical density is given in Tables 13 and 14. The characteristic feature
of the effect of the BR scaling is summarized in Figs.11, 12 and 13.
The remarkable thing to notice is that the critical density lowered to
$u_c\sim 2$ is completely insensitive to the parameters such as the
form of the symmetry energy, the constants $C_{\Lambda^\star}^{S,T}$ etc. in which
possible uncertainties of the theory lie.  

In order to verify the key hypothesis of the BR scaling -- that the scaling
subsumes higher order effects, thus {\it suppressing} higher chiral order 
effects of the scaled Lagrangian,  we keep all the parameters 
{\it fixed} at the values determined at $\O (Q^3)$, BR-scale as described above,
then ignore all terms of $\O (Q^3)$ ({\ie}, loop corrections) and
(A) set $C_{\Lambda^\star}^S=C_{\Lambda^\star}^T=0$ and (B) 
set $(C_{\Lambda^\star}^S-C_{\Lambda^\star}^T)f^2
=10$ and calculate the critical density. The results are given in Table 15.
We see that the effects of loop corrections and four-Fermi interactions
amount to less than 10\%. This may be taken as an {\it \`a posteriori}
verification of the validity of the assumption made in deriving the scaling,
although it is difficult to assess how reliable the absolute value of the 
predicted critical density (and the corresponding equation of state of the
condensed phase) is.


\section{Discussion}
\indent

In conclusion, we have shown that chiral perturbation theory at
order N$^2$LO predicts kaon condensation in ``nuclear
star" matter at a density $2\lsim u_c \lsim 4 $ with
a large fraction of protons -- $x=0.1\sim 0.2$ at the critical point and
rapidly increasing afterwards -- neutralizing the negative charge of
the condensed kaons. For this to occur, four-Fermi interactions
involving $\Lambda^\star$ are found to play an important role in driving
the
condensation but the critical density is negligibly dependent on the
strength of the four-Fermi interaction. 

It is found that the BR scaling \cite{br94,br91}
favors a condensation at a density as low as twice the matter density and
that when the BR scaling is operative, higher chiral corrections coming from
the scaled Lagrangian are insignificant, justifying the basic assumption
that goes into the derivation of the scaling relation. This suggest that
at least for kaon condensation, the tree approximation with the scaled
Lagrangian is consistent with the basic idea of the BR scaling.

Given the relatively low critical density obtained in the higher-order
calculation of this paper, we consider it reasonable to assume that
the compact-star properties will be qualitatively the same as in the
tree-order calculation of ref. \cite{TPL}. We will report on this matter in 
a future publication.

Our treatment is still far from self-consistent as there are many nuclear 
correlation effects that are still to be
taken into account. How to incorporate them in full consistency 
with chiral symmetry is not known. In particular the role of four-Fermi
interactions in the non-strange channel, {\eg}., short-range nuclear
correlations which involve both interaction terms in the Lagrangian and
nuclear many-body effects, is still poorly understood.
A problem of this sort may have to be  
addressed in terms of renormalization group flows as in condensed matter
physics \cite{fixedpoint}. A work is in progress along this line \cite{hkl}.

\vskip 5mm

\centerline{\large \bf Acknowledgments}
\vskip 0.3cm
We acknowledge useful discussions with H. Jung, K. Kubodera, A. Manohar,
F. Myhrer, V. Thorsson, A. Wirzba, W. Weise and H. Yabu.
The work of CHL and DPM were supported in part by the Korea Science and
Engineering Foundation through the CTP of SNU and in part by 
the Korea Ministry of Education under Grant No. BSRI-94-2418 
and the work of GEB by the US Department of Energy under Grant
No. DE-FG02-88ER40388.

\newpage


\def\pumu{\partial^\mu}
\def\pdmu{\partial_\mu}

\def\barB {\bar{B}}
\def\barN {\bar{N}}
\def\dagN {N^\dagger}
\def\barK {\bar{K}}
\def\L{{\cal L}}
\def\lrarrow {\stackrel{\leftrightarrow}{{\partial}}_\mu}
\def\lrarrowu{\stackrel{\leftrightarrow}{{\partial^\mu}}}
\def\Mii{(\frac{1}{\sqrt{2}}\pi^0+\frac{1}{\sqrt{6}}\eta)}
\def\Mjj{(-\frac{1}{\sqrt{2}}\pi^0+\frac{1}{\sqrt{6}}\eta)}
\def\Mkk{(-\sqrt{\frac{2}{3}}\eta)}

\def\D{D}
\def\F{F}

\section*{Appendix A: Vertices and Feynman Rules }
\renewcommand{\theequation}{A.\arabic{equation}}
\setcounter{equation}{0}
\indent

In this Appendix, the chiral Lagrangian used is given explicitly in
terms of the component fields (meson octet, baryon octet and decuplet)
we are interested in. The vertices of the
Feynman graphs can be read off directly.
The subscript $n$ in ${\cal L}_n$ denotes the number of lines attached to the
given vertex. In Fig.1, the three-point vertex ${\cal L}_3$ enters in the
diagrams (c,d,f), the four-point vertices ${\cal L}_4^A$ (4-meson) and 
${\cal L}_4^B$(2-meson + 2-baryon) in the diagrams 
(b,d) and (b,c,e), respectively and the higher-point vertices 
${\cal L}_5$ and ${\cal L}_6$ figure in the diagrams (f) and (a), respectively.
\be
\L_3 &=&- \frac{1}{f} \left[
     \bar p_v S_v^\mu p_v \left( (\D+\F) 
    \partial_\mu \pi^0 - 
  \frac{1}{\sqrt{3}} (\D-3\F)\partial_\mu \eta \right)\right.\nonumber\\
&&  \;\;\;\;\;\;\;\;\;\;
  \left. +  \bar n_v S_v^\mu n_v \left(-( \D+\F) 
    \partial_\mu\pi^0
  -\frac{1}{\sqrt{3}} (\D-3\F)\partial_\mu \eta \right)\right]
 \nonumber\\ 
 &&- \left. \frac{1}{f} \right[
     \sqrt 2 (\D+\F) \bar p_v S_v^\mu n_v \partial_\mu\pi^+
 +\sqrt 2 (\D-\F) \bar p_v S_v^\mu \Sigma^+_v \partial_\mu K^0
   +\sqrt 2 (\D-\F) \bar n_v S_v^\mu \Sigma^-_v \partial_\mu K^+ 
\nonumber\\ && \;\;\;\;\;\;\;\;\;\;
 -\frac{1}{\sqrt{3}}(\D+3\F)  \bar p_v  S_v^\mu \Lambda_v \partial_\mu K^+
  -\frac{1}{\sqrt{3}}(\D+3\F) \bar n_v  S_v^\mu \Lambda_v \partial_\mu K^0
\nonumber\\ && \;\;\;\;\;\;\;\;\;\;
\left.  +(\D-\F) \bar p_v S_v^\mu \Sigma^0_v \partial_\mu K^+
  -(\D-\F) \bar n_v S_v^\mu \Sigma^0_v \partial_\mu K^0 
  +h.c.
\right]
\nonumber \\
&&-\frac{C}{\sqrt{6}f}\left[ \sqrt{2}\bar p_v \partial^\mu\pi^0
\Delta^+_{v,\mu}+\bar p_v\partial^\mu\pi^+\Delta^0_{v,\mu}
+\frac{1}{\sqrt{2}}\bar p_v\partial^\mu K^+ \Sigma^{*0}_{v,\mu}
-\sqrt 3\bar p_v \partial^\mu\pi^-\Delta^{++}_{v,\mu}
\right. \nonumber\\ &&\;\;\;\;\;\;\;\;\;\;  
-\bar p_v \partial^\mu K^0 \Sigma^{*+}_{v,\mu} 
+ \sqrt{ 2}\bar n_v \partial^\mu\pi^0
\Delta^0_{v,\mu}-\bar n_v \partial^\mu\pi^-\Delta^+_{v,\mu}
-\frac{1}{\sqrt{2}}\bar n_v \partial^\mu K^0 \Sigma^{*0}_{v,\mu}
 \nonumber\\
  &&\;\;\;\;\;\;\;\;\;\;  \left. 
+\sqrt 3 \bar n_v \partial^\mu\pi^+\Delta^-_{v,\mu}
+\bar n_v \partial^\mu K^+ \Sigma^{*-}_{v,\mu} 
\vphantom{\frac 12} +h.c. \right] 
\\ 
&& \vphantom{\Sigma^+} \nonumber\\
\L_{4}^A &=& 
   -\frac{1}{6f^2}\left[ \frac 12 
   \left((\pi^0+\sqrt{3}\eta)\lrarrow K^+\right)
   \left((\pi^0+\sqrt{3}\eta)
   \lrarrowu K^-\right) \right. \nonumber\\
&& \;\;\;\;\;\;\;\;\;\;
   +\left(-(K^+\lrarrow K^-)(\pi^+\lrarrowu \pi^-)
       +(K^+\lrarrow\pi^-)(K^-\lrarrowu\pi^+) \right)\nonumber\\
&& \;\;\;\;\;\;\;\;\;\;
   - (K^+\lrarrow K^-)(K^+ \lrarrowu K^-) \nonumber\\
&& \;\;\;\;\;\;\;\;\;\;
   \left. +\left((K^+\lrarrow K^-)(\bar K^0\lrarrowu K^0) +
       (K^+\lrarrow \bar K^0)(K^-\lrarrowu K^0)\right) \right] \nonumber\\
 & & +  \frac{B_0}{6 f^2} \left[  K^+ K^-  \left(
   \frac{1}{2}(3m_u+m_s)(\pi^0)^2 +\frac{1}{\sqrt{3}}(m_u-m_s)\pi^0\eta
    +\frac{1}{2}(m_u+3m_s)\eta^2 
              \right) \right. \nonumber \\
  &&   \;\;\;\;\;\;\;\;\;\; +  ( 2 m_u +m_d + m_s) \pi^+\pi^-   K^+ K^-   
     +  (  m_u +m_s) K^+ K^-  K^+ K^- \nonumber  \\
 &&\;\;\;\;\;\;\;\;\;\; + \left. \vphantom{\frac{B_0}{6 f^2}}  
  ( m_u + m_d + 2 m_s)    K^+  K^-  \bar K^0   K^0  \right]
\\
&& \vphantom{\Sigma^+} \nonumber\\
\L_4^{B} &=&
 -\frac{ i}{4f^2} \left[\vphantom{\frac 12}
 (K^+\lrarrow K^-)\left( 2  \bar \Xi^-_v v^\mu \Xi^-_v 
 -\bar \Sigma^+_v v^\mu \Sigma^+_v + \bar\Sigma^-_v v^\mu \Sigma^-_v
 +\bar\Xi^0_v v^\mu\Xi^0_v -2\bar p_v v^\mu  p_v -\bar n_v v^\mu n_v\right)
 \right.
\nonumber\\
&& \;\;\;\;\;
  - \bar p_v  v^\mu  p_v \left\{ \left( \pi^+\lrarrow\pi^-\right) -
  \left( \bar K^0\lrarrow K^0\right) \right\}
   +\bar n_v  v^\mu  n_v \left\{\left(\pi^+\lrarrow\pi^-\right)
   -2 \left( K^0\lrarrow\bar K^0\right)
   \right\}\left.\vphantom{\frac 12}\right]
\nonumber\\
 && \left. + \frac{i}{4f^2} \right[
        \bar p_v v^\mu  n_v \left( K^+ \lrarrow \bar K^0\right)
       - \bar \Sigma^+_v  v^\mu  p_v \left( K^-\lrarrow \pi^+\right)
    -\frac{1}{\sqrt{2}} \bar n_v  v^\mu  \Sigma^-_v 
 \left(  (\pi^0 +\sqrt{3}\eta)\lrarrow K^+ \right)
 \nonumber\\ && \;\;\;\;\;\;\; 
       + \frac{1}{\sqrt{2}}\bar n_v  v^\mu  \left(\Sigma^0_v -\sqrt{3}
      \Lambda_v \right) \left(\pi^-\lrarrow K^+\right)
       - \frac 12 \bar p_v  v^\mu  \left(\Sigma^0_v+\sqrt{3}
       \Lambda_v\right)\left( (\pi^0 +\sqrt{3}\eta)\lrarrow K^+ \right)
 \nonumber\\ && \;\;\;\;\;\;\; \left.
\vphantom{\frac 12} +h.c. \right]
 \nonumber\\
 && -i \frac{3}{4f^2} \left[ \left(K^+ v\cdot 
\stackrel{\leftrightarrow}{{\partial}} K^-\right) 
 \left( \bar \Delta^{++}_{v,\nu} \Delta^{++,\nu}_v
  +\frac{2}{\sqrt 3} \bar\Delta^+_{v,\nu}
\Delta^{+,\nu}_v +\frac {1}{\sqrt 3} \bar\Sigma^{*+}_{v,\nu} 
 \Sigma^{*+,\nu}_v 
+\frac{1}{\sqrt 3} \bar\Delta^0_{v,\nu} \Delta^{0,\nu}_v \right.  \right.
\nonumber\\ && \left. \;\;\;\;\;\;\;\;\;\;
\left.
-\frac{1}{\sqrt 3}\bar\Xi^{*0}_{v,\nu}
\Xi^{*0,\nu}_v -\frac{1}{\sqrt 3}\bar\Sigma^{*-}_{v,\nu} \Sigma^{*-,\nu}_v
-\bar\Omega^-_{v,\nu}\Omega^{-,\nu}_v
-\frac{2}{\sqrt 3}\Xi^{*-}_{v,\nu} \Xi^{*-,\nu}_v \right)\right]
 \\
&& \vphantom{\Sigma^+} \nonumber\\
\L_5 &=& \left. \frac{1}{6\sqrt{2}f^3} \right[ K^+ K^- 
  \left\{ \vphantom{\frac 12}
   \sqrt 2  \bar p_v  S_v^\mu p_v 2 \F 
(\partial_\mu\pi^0+ \sqrt{3}\partial_\mu\eta) 
     - \sqrt 2 \bar n_v  S_v^\mu n_v (\D -\F )
(\partial_\mu\pi^0+ \sqrt{3}\partial_\mu\eta) \right\}
 \nonumber\\  && \;\;\;\;\;\;\;\;\;\;
  + K^+K^- \left\{ \vphantom{\frac 12} \right.
      (\D+\F) \bar p_v  S_v^\mu n_v \partial_\mu\pi^+
 + (\D-\F) \bar p_v S_v^\mu \Sigma^+_v \partial_\mu K^0
\nonumber\\ && \;\;\;\;\;\;\;\;\;\;\;\;\;\;\;\;\;\;\;\;
   + 2 (\D-\F) \bar n_v S_v^\mu \Sigma^-_v \partial_\mu K^+
     -\frac{2}{\sqrt{6}}(\D+3\F)  \bar p_v   S_v^\mu \Lambda_v 
        \partial_\mu K^+
\nonumber\\ && \;\;\;\;\;\;\;\;\;\;\;\;\;\;\;\;\;\;\;\;
  -\frac{1}{\sqrt{6}}(\D+3\F) \bar n_v   S_v^\mu \Lambda_v 
  \partial_\mu K^0
 +\frac{2}{\sqrt{2}}(\D-\F) \bar p_v S_v^\mu \Sigma^0_v \partial_\mu K^+ 
\nonumber\\ &&\left.\left. \;\;\;\;\;\;\;\;\;\;\;\;\;\;\;\;\;\;\;\;
  -\frac{1}{\sqrt{2}}(\D-\F) \bar n_v S_v^\mu \Sigma^0_v \partial_\mu K^0 
  + h.c.
\right\}\right]
\nonumber \\
&&+\frac{C}{12\sqrt{2} f^3}
  \left[ \vphantom{\frac 12}\right.
 ( K^+ K^-)\left(\sqrt{\frac 23}\bar p_v\partial^\mu\pi^0
\Delta^+_{v,\mu} +\sqrt{2}\bar p_v
\partial^\mu\eta\Delta^+_{v,\mu}+\frac{1}{\sqrt{3}}
\bar p_v \partial^\mu\pi^+\Delta^0_{v,\mu} \right.
\nonumber\\ && \;\;\;\;\; \;\;\;\;\;\;\;\;\;\; 
 +\sqrt{\frac23}\bar p_v \partial^\mu K^+ \Sigma^{*0}_{v,\mu }
 -\bar p_v \partial^\mu\pi^-\Delta^{++}_{v,\mu}-\frac{1}{\sqrt{3}}
\bar p_v \partial^\mu K^0\Sigma^{*+}_{v,\mu}
+\sqrt{\frac 23}\bar n_v \partial^\mu\pi^0\Delta^0_{v,\mu} 
\nonumber\\ && \;\;\;\;\; \;\;\;\;\;\;\;\;\;\; 
+ \sqrt{2} \bar n_v \partial^\mu\eta \Delta^0_{v,\mu }
+ \bar n_v \partial^\mu\pi^+ \Delta^-_{v,\mu  }
+\frac{2}{\sqrt{3}}\bar n_v \partial^\mu K^+\Sigma^{*-}_{v,\mu}
\nonumber\\
&& \;\;\;\;\; \;\;\;\;\;\;\;\;\;\; \left. 
-\frac{1}{\sqrt{3}}\bar n_v \partial^\mu\pi^-\Delta^+_{v,\mu}
 -\frac{1}{\sqrt{6}} \bar n_v \partial^\mu K^0 \Sigma^{*0}_{v,\mu}\right) 
  \left. \vphantom{\frac 12} + h.c. \right]
\\
&& \vphantom{\Sigma^+} \nonumber\\
\L_6 & =& -\frac{i}{96 f^4} \left[ \vphantom{\frac 12}\right.
	 \bar p_v  v^\mu p_v  \left\{ \vphantom{\frac 12}\right. 
          (K^+\lrarrow K^-)( 5\pi^+ \pi^- +5 \bar K^0 K^0
+8K^+K^- +(\pi^0)^2 +2 \sqrt{3}\pi^0\eta +3 \eta^2)  \nonumber\\
&&\left.\;\;\;\;\;\;\;\;\;\;  \;\;\;\;\;\;\;\;\;\;  
   + 7 K^+ K^-( \pi^+\lrarrow\pi^- ) -7(\bar K^0 \lrarrow K^0) K^+ K^-
\vphantom{\frac12}  \right\}
\nonumber\\
 &&  \;\;\;\;\;\;\;\;\;\;\; 
    + \bar n_v  v^\mu n_v\left\{\vphantom{\frac12}\right. (K^+\lrarrow K^-)
    \left( \frac{(\pi^0)^2}{2} +\sqrt{3}\pi^0\eta +\frac{3}{2}\eta^2 -2\pi^+\pi^-
   +4 K^+ K^-+7\bar K^0 K^0 \right)  \nonumber \\
&& \left. \vphantom{\frac 12}\;\;\;\;\;\;\;\;\;\; \;\;\;\;\;\;\;\;\;\; 
  -5 (\bar K^0\lrarrow K^0) K^+ K^- 
   +2 K^+ K^- (\pi^+\lrarrow\pi^-) \left.\vphantom{\frac 12} \right\}\right]
\label{kn6}\ee
where the quark masses are related to the meson masses as
\be
M_\pi^2 = B_0 (m_u+m_d) = 2 B_0 \hat m,\;\;
M_\eta^2 = \frac 23 B_0 (\hat m+ 2 m_s),\;\;
M_K^2 = B_0 (\hat m+m_s).
\ee

The propagator for the baryon octet in HFF is
$i/(v\cdot k)$, where $k^\mu$ is the residual momentum of the baryon as
defined by $p^\mu =m_B v^\mu +k^\mu$ \cite{JM}. The propagator for the decuplet
is also simplified to the form
 $[i/(v\cdot k-\delta_T)] \times (v^\mu v^\nu -g^{\mu\nu}
  -4\frac{d-3}{d-1} S_v^\mu S_v^\nu)$. In the HFF, 
the spin operator takes the form
$S^\mu_v S^\nu_v =\frac 14 (v^\mu v^\nu-g^{\mu\nu})
  +\frac i2 \epsilon^{\mu\nu\alpha\beta} v_\alpha S_{v,\beta}$.

Given the vertices and the propagators, one can immediately write down the
the integral entering in the diagram (e) of Fig.1,
\be
I^E_i &=& 
 \frac 1i \int \frac{d^n k}{(2\pi)^n} \frac{v\cdot (2q-k)}{v\cdot k+i\epsilon}
 \frac{v\cdot (q+q^\prime-k)}{(q-k)^2-M_i^2+i\epsilon}
\nonumber\\
&=& -(v\cdot q^\prime +2 v\cdot q)\Delta_i +2 v\cdot q v\cdot (q+q^\prime)
\Sigma_i (v\cdot q)
\ee
where
\be
\Delta_i &=& \frac 1i\int\frac{d^nk}{(2\pi)^n} \mu^\epsilon
\frac{1}{k^2-M_i^2+i\epsilon}
= -\frac{M_i^2}{16\pi^2}\ln\frac{M_i^2}{\mu^2} -M_i^2 2 L
\nonumber\\
\Sigma_i(\omega) &=& \frac 1i \int\frac{d^nk}{(2\pi)^n}
\frac{1}{k^2-M_i^2+i\epsilon} \frac{1}{v\cdot k+\omega+i\epsilon}
\nonumber\\
&=& \frac{1}{8\pi^2} \left[
\omega\left(1-\ln\frac{M_i^2}{\mu^2}\right)+\bar f_i(\omega)\right] -4\omega L.
\ee
Here the divergent term $L$ and the finite loop term
$\bar f_i(\omega)$ are given by
\be
L &=& \frac{1}{16\pi^2} \mu^{-\epsilon}\left(
 -\frac 2\epsilon +\gamma-1-\ln 4\pi\right)
\nonumber\\
\bar f_i(\omega) &=& \sqrt{\omega^2-M_i^2}
\left( \ln\left|
\frac{\omega-\sqrt{\omega^2-M_i^2}}
     {\omega +\sqrt{\omega^2-M_i^2}}
\right| +i 2\pi\theta(\omega-M_i) 
 \right)  \;\;{\rm for}\;\; |\omega| > M_i
\nonumber\\
&=& -\sqrt{M_i^2-\omega^2}\left(\pi + 2\tan^{-1}
 \frac{\omega}{\sqrt{M_i^2-\omega^2}}
\right)  \;\;\;\;\;\;{\rm for}\;\; |\omega| < M_i.
\ee
Note that $\bar f_i(\omega)$ has a kink at $\omega=M_i$, which
explains the nontrivial behavior of the $KN$ scattering amplitude seen in
Fig.2.
In order to get one-loop results, we also need the following well-known
integrals
\be
J_i(q^2) &=& \frac 1i\int\frac{d^nk}{(2\pi)^n}  \mu^\epsilon 
\frac{1}{k^2-M_i^2+i\epsilon}\frac{1}{(k-q)^2-M_i^2+i\epsilon}
\nonumber\\
 &=&  -\frac{1}{16\pi^2}\left(1+\ln\frac{M_i^2}{\mu^2}\right) -2L
 +\bar J_i(q^2)
\ee
where 
\be
\bar J_i(t) &=& \frac{1}{16\pi^2}\left[2-\sqrt{1-4M_i^2/t}
\; \ln\frac{\sqrt{1-4M_i^2/t}+1}{\sqrt{1-4M_i^2/t}-1}
 \right] \;.
\ee
All the integrals needed for the diagrams in Fig.1 
can be obtained from $\Delta_i$, $J_i(q^2)$ and $\Sigma_i(\omega)$.


\section*{Appendix B: Mass and Wave Function Renormalizations}
\renewcommand{\theequation}{B.\arabic{equation}}
\setcounter{equation}{0}
\indent

At one-loop order, we need one counter term, $ {\cal L}_{c} = 
-\delta m_B \Tr \bar B B$, to renormalize the nucleon mass. It does not
affect $KN$ scattering amplitude, but it is needed to 
absorb the divergences coming from one-loop self-energy graphs.
The nucleon self-energy including the counter terms is
   \be
   \Sigma_N(\omega) &=& \delta m_B -2 a_1 \hat m -2 a_2 m_s -2 a_3 (m_s+2\hat m)
    -2 \omega \bar h -c_1 \omega^3
   \nonumber\\
   && +\sum_i \frac{\lambda_i}{f^2} {\cal H}_i(\omega)
   +\sum_i \frac 43 \frac{\lambda_{D,i}}{f^2} {\cal H}_i(\omega-\delta_T)
   \ee
where $\omega=v\cdot p -m_B v$ and
   \be
   \bar h &=& h_1 \hat m+h_2 m_s +h_3 (2\hat m +m_s) \nonumber\\
   {\cal H}_i (\omega) &=& -\frac 14 \left[ \omega \Delta_i
    +(M_i^2-\omega^2) \Sigma_i(\omega) \right]
   \nonumber\\
   \Delta_i &=& -\frac{M_i^2}{16\pi^2} \ln\frac{M_i^2}{\mudr^2}
    -M_i^2 2 L
   \nonumber\\
   \Sigma_i(\omega) &=& \frac{1}{8\pi^2}\left[\omega \left(1
   -\ln\frac{ M_i^2}{\mudr^2}\right) - f_i(\omega) \right] -4\omega L
   \nonumber\\
   f_i(\omega) &=& \sqrt{M_i^2 -\omega^2}\left(\pi  + 2\sin^{-1}
   \frac{\omega}{M_i} \right)
   \nonumber\\
   L&=& \frac{1}{16\pi^2} \mudr^{-\epsilon}\left( -\frac{2}{\epsilon}+
   \gamma-1-\ln 4\pi \right)
   \ee
and the coefficients $\lambda_i$'s are
   \be
   \lambda_\pi &=& 3(D+F)^2 \nonumber\\
   \lambda_K &=& \frac 13 \left( (D+3F)^2 + 9 (D-F)^2\right) \nonumber\\
   \lambda_\eta &=& \frac 13 (D-3F)^2 \nonumber\\
   \lambda_{D,\pi} &=& 4\lambda_{D,K} = 2 C^2.
   \ee
Here $\mudr$ stands for the arbitrary mass scale $\mu$ 
that arises in the dimensional regularization. 
One can compute $Z_0$ and $Z_3$ from $\Sigma_N(\omega)$
or write down the corresponding counter terms as in renormalizable theory,
$(Z_0 -1) m_B \Tr \bar B B $
 and $(Z_3 -1) \Tr\bar B iv\cdot\partial B$.\footnote{In standard notation,
$Z_0$ ($Z_3$) corresponds to the mass (wave function) renormalization.
The counter term ($Z_0-1) m_B\Tr\bar B B$ corresponds to $\delta m_B\Tr\bar
B B$, and $(Z_3-1) \Tr\bar B iv\cdot\partial B$ to the
$h_i$ terms in the $\nu=3$ Lagrangian. }
However in a non-renormalizable theory such as ours, many counter terms enter
at order by order, so it is more convenient to use the constraints on 
$\Sigma_N(\omega)$ directly.  Using a physical scheme,
the convenient constraints are 
   \be
   \left. \Sigma_N(\omega)\right|_{\omega=0} = 0, \;\;\;\;\; 
   \left. \frac{\partial\Sigma_N(\omega)}{\partial \omega} \right|_{\omega=0}
    = 0.
   \ee
In this case, the $m_B$, that figures in the leading-order Lagrangian,
can be identified as the physical baryon mass.
The constant $\delta m_B$ is determined by the first constraint 
to absorb the
divergence and $\bar h$, relevant for $KN$ scattering, can be fixed by the 
second constraint. The latter is explicitly given by
   \be
   \left. \frac{\partial \Sigma_N (\omega)}{\partial\omega} \right|_{\omega=0}
   &=& - 2 \bar h +\sum_i \frac{\lambda_i}{f^2}
    \left( -\frac 14 \Delta_i + M_i^2\left. \frac{\partial \Sigma_i (\omega)}
      {\partial\omega}\right|_{\omega=0} \right)
   \nonumber\\
   &&  +\sum_i \frac 43 \frac{\lambda_{D,i}}{f^2}
    \left( -\frac 14 \Delta_i + 2\delta_T\Sigma(-\delta_T) + (M_i^2-\delta_T^2)
    \left. \frac{\partial \Sigma_i (\omega)}
   {\partial\omega}\right|_{\omega=-\delta_T} \right).
   \label{eq7}\ee
The constraints
   \be
   Z_N-1 =\left. \frac{\partial \Sigma_N (\omega)}{\partial\omega} 
   \right|_{\omega=0} =0
   \label{eq8}\ee
completely determine $\bar h$ as a function of $\mudr$.
One convenient choice is $ \bar h =\bar h^r(\mudr) + \sum_i \alpha_i L $
where $L$ is the divergent piece, and $\bar h^r$ is the finite part
that includes the chiral log terms $\ln (M_i^2/\mudr^2)$. 
The renormalized parameter $\bar h^r(\mudr)$ is related with each other
at different scales $\mudr$ through the relation
   \be
   \frac{\bar h^r(\mudr =\mu_1)}{\bar h^r(\mudr=\mu_2)}
    =\sum\beta_i(\delta_T) \ln\frac{\mu_1}{\mu_2}\;\; .
   \ee

Consider now kaon self-energy which we can write as
   \be
   \Pi_K (q^2) &=& \sum_i \frac{\alpha_i}{12f^2}\Delta_i + c.t.
   \ee
where $\alpha_i $ are given by
   \be
   \alpha_\pi = -3 M_\pi^2,\;\;\;\;
   \alpha_K = -6 M_K^2,\;\;\;\;
   \alpha_\eta = M_\pi^2 .
   \ee
In the physical scheme, the constraints are 
   \be
   \left. \Pi_K(q^2)\right|_{q^2=M_K^2} = 0 ,\;\;\;\;
   Z_K-1= \left. \frac{\partial \Pi_K(q^2)}{\partial q^2}\right|_{q^2=M_K^2} = 0
   \ee
where $M_K$ is the physical kaon mass.\footnote{Here we pick the
renormalization point at $\mu_K=M_K$ for the kaon wave function renormalization.
For $\pi$ and $\eta$,  we take
$M_\pi$ and $M_\eta$ in the tree-order Lagrangian
as physical masses. 
This physical scheme is independent of the arbitrary mass scale $\mudr$ 
figuring in the dimensional regularization.} 
Note however that the counter terms in kaon self-energy do not affect 
the $KN$ scattering directly. So we shall not need
the explicit magnitudes of these counter terms for our purpose.

Finally the scattering amplitudes can be obtained simply by calculating
the six topologically distinct diagrams of Fig. 1 using physical masses
   \be {\cal T}^{KN} =\sum_{i=a,...,f} {\cal T}_i^{KN} .
   \ee
This is 
because there is no contribution from the mass and wave function 
renormalizations,
   \be
   {\cal T}^{KN}_{w.r} = \alpha \frac{(1-Z_N+1-Z_K)}{2f^2} v\cdot(q+q^\prime)
   \ee
where $\alpha=2$ for $K^+p$ and $\alpha=1$ for $K^+n$.


\renewcommand{\theequation}{C.\arabic{equation}}
\setcounter{equation}{0}
\section*{Appendix C: Renormalization of ${\cal O}(Q^3)$ Counter Terms }
\indent

The quantity $\bar g_s (\bar g_v)$ is the crossing-odd t-channel isoscalar
(isovector) contribution from one-loop plus counter terms which after the
standard dimensional regularization, takes the form
   \be
   \bar g_{s,v} = \alpha_{s,v} +\beta_{s,v} + \frac{1}{32\pi^2} \frac{1}{f^2 M_K^2}
    \gamma_{s,v}+ \sum_{i=\pi,K,\eta} \delta_{{s,v}}^i\frac{1}{f^2 M_K^2}
    \left(L+\frac{1}{32\pi^2} \ln \frac{M_i^2}{\mu^2} \right)
   \label{eq10}\ee
where $L$ contains the divergence, and  
   \be
   \alpha_s &=&-\frac{1}{M_K^2}\left(-\frac 34\bar h+(l_1-2 l_2 +l_3)
  (\hat m+m_s) \right) 
   \nonumber\\
   \beta_s &=&-\frac 12\left((g_2-2 g_3+g_6)+(g_4-2 g_5+g_7)+(g_8-2 g_9)\right)
   \nonumber\\
   \alpha_v &=& -\frac{1}{M_K^2} \left(-\frac 14\bar h +(l_1+l_3)(\hat m+m_s)
    \right) 
   \nonumber\\
   \beta_v &=& - \frac 12\left( (g_2 +g_6) +(g_4+g_7) +g_8\right)
\ee
with
\be
   \bar h &=& h_1 \hat m +h_2 m_s +h_3 (2\hat m +m_s).
   \ee
Here we have separated $\alpha_{s,v}$ and $\beta_{s,v}$ because in off-shell
amplitudes, the constant $\alpha_{s,v}$ is multiplied by
$\omega$ while the $\beta_{s,v}$ is multiplied by $\omega^3$,
thus behaving differently for $\omega\neq M_K$.  The constants
$\gamma_{s,v}$ -- coming from finite loop terms -- and $\delta^{s,v}$ --
multiplying the divergence -- are given by
   \be
   \gamma_s &=& -3 M_K^2 -\frac 94 M_K f_\pi (-M_K) -\frac 94 M_K f_\eta(-M_K)
      -\frac 94 (D+F)^2 M_\pi^2 
   \nonumber\\
   &&-\frac 14 (D-3 F)^2 M_\eta^2 +\left( \frac 32 (D-F)^2 +\frac 16 (D+3F)^2
    \right) M_K^2
   \nonumber\\
   && +|C|^2 \left( \left(1+\sqrt 3\right) C_\pi (-\delta_T)
     +\frac{1}{3\sqrt 3} C_K (-\delta_T) -\frac 14 F_K (-\delta_T) \right)
   \nonumber\\
   \gamma_v &=& \frac 13 M_\pi^2 -\frac 43  M_K^2 +\frac 14 M_K f_\pi (-M_K)
      -\frac 34 M_K f_\eta(-M_K)
      +\frac {11}{12} (D+F)^2 M_\pi^2 
   \nonumber\\
   &&-\frac 1{12} (D-3 F)^2 M_\eta^2 +\left( - \frac 76 (D-F)^2 +
     \frac 1{18} (D+3F)^2 \right) M_K^2
   \nonumber\\
   && +|C|^2 \left( \left(1+\frac{1}{3\sqrt 3}\right) C_\pi (-\delta_T)
     +\frac{1}{\sqrt 3} C_K (-\delta_T) +\frac 29 F_\pi (-\delta_T)
     +\frac{1}{36} F_K(-\delta_T) \right)
   \nonumber\\
   \delta_{s}^{\pi} &=& \frac  94 M_K^2 -\frac 34 M_\pi^2 
       -\frac{27}{8} (D+F)^2 M_\pi^2 +|C|^2 \left( 3+3\sqrt 3\right) 
      (2\delta_T^2 -M_\pi^2 )
   \nonumber\\
   \delta_{s}^{K} &=& -\frac 32 M_K^2 -\left(\frac{27}{8} (D-F)^2 +\frac 38
     (D+3 F)^2 \right) M_K^2 +|C|^2\left(\frac 34+\frac{1}{2\sqrt 3}\right)
     (2\delta_T^2 -M_K^2 )
   \nonumber\\
   \delta_{s}^{\eta} &=& \frac 94 M_K^2 -\frac 34 M_\eta^2 -\frac 38(D-3F)^2
    M_\eta^2
   \nonumber\\
   \delta_{v}^{\pi} &=& -\frac{1}{4} M_K^2 -\frac 14 M_\pi^2 -\frac 98 (D+F)^2
     M_\pi^2 +|C|^2 \left(\frac 53 +\frac{1}{\sqrt 3} \right) (2\delta_T^2-M_\pi^2)
   \nonumber\\
   \delta_{v}^{K} &=& \frac 12 M_K^2 -\left(\frac 98 (D-F)^2 +\frac 18 (D+3F)^2
   \right) M_K^2 +|C|^2\left(-\frac{1}{12}+\frac{\sqrt 3}{2} \right)
   (2\delta_T^2-M_K^2)
   \nonumber\\
   \delta_{v}^{\eta} &=&\frac 34 M_K^2 -\frac 14 M_\eta^2 -\frac 18 (D-3F)^2
   M_\eta^2
   \ee
where $C_i(-\delta_T), F_i(-\delta_T)$ and $f_i(\omega)$ are
   \be
   C_i(-\delta_T) &=& -(M_i^2 +\delta_T^2) -3\delta_T f_i(-\delta_T)
   \nonumber\\
   F_i(-\delta_T) &=& -\frac 43 M_i^2 +6\delta_T^2 +6\delta_T f_i(-\delta_T)
   \nonumber\\
   f_i(\omega) &=& 
       \sqrt{\omega^2-M_i^2}
       \ln\left|\frac{\omega+\sqrt{\omega^2-M_i^2}}{\omega-\sqrt{\omega^2-M_i^2}}
     \right| \;\;\;\;\; {\rm for}\;\; \omega^2 > M_i^2
   \nonumber\\
   &=& \sqrt{M_i^2-\omega^2}\left(\pi +2 \sin^{-1}\frac{\omega}{M_i}\right)
      \;\;\;\;\; {\rm for}\;\; \omega^2 < M_i^2  .
   \ee
Note that $f_i(\omega)$ result from diagram of Fig. 1e as explained in Appendix A.
We write the counter terms as
   \be
   \alpha_{s,v} = \alpha_{s,v}^r + \alpha_{s,v}^{div} ,\;\;\;\;\;
   \beta_{s,v} = \beta_{s,v}^r + \beta_{s,v}^{div}
   \ee
where the divergent parts
$\alpha_{s,v}^{div}$ and $\beta_{s,v}^{div}$ are to 
cancel the divergence part  $L$ in eq.(\ref{eq10}) and the finite counter
terms $\alpha^r$ and $\beta^r$ are to be fixed by experiments.
After removing the divergences, the constant $\bar g_{s,v}$ 
can be written as
   \be
   \bar g_{s,v} &=& \alpha_{s,v}^r +\beta_{s,v}^r 
   + \frac{1}{32\pi^2} \frac{1}{f^2 M_K^2}
   \left( \gamma_{s,v}+ \sum_{i=\pi,K,\eta} \delta_{{s,v}}^i
   \ln \frac{M_i^2}{\mu^2}\right) .
   \ee
Note that $\bar g_{s,v} $ is $\mu$-independent and hence can
be fixed from experiments independently of $\mu$. How to do this for off-shell
amplitudes is described in the main text.


\renewcommand{\theequation}{D.\arabic{equation}}
\setcounter{equation}{0}
\section*{Appendix D: $1/m_B$ Corrections }
\indent

In writing down the off-shell amplitudes with the parameters determined
on-shell, we have assumed that the $\O (Q^2)$ terms quadratic in $\omega$
can be calculated by saturating the Born graphs with the octet and decuplet
intermediate states. In calculating the decuplet
contributions in the lowest order in $1/m_B$ from the Feynman graphs in
relativistic formulation, one encounters the usual
off-shell non-uniqueness characterized by a factor Z in the 
decuplet-nucleon-meson vertex when the decuplet is off-shell:
   \be
   {\cal L} &=& C \bar T_\mu \left( g^{\mu\nu} -\left(Z+\frac 12 \right)\gamma^\mu
   \gamma^\nu\right) A_\nu B + h.c. \; .
   \ee
This gives the $Z$ dependence in the $1/m_B$ corrections to $\bar d_{s,v}$: 
   \be
   \bar d_{s,\frac 1m} &=& -\frac{1}{48}\left( (D+3F)^2 +9 (D-F)^2\right)
     \frac{1}{m_B} -\frac{1}{12}|C|^2\frac{1}{m_B} \left(Z+\frac 5 2\right)
     \left(\frac 12 -Z\right)
   \nonumber\\
   \bar d_{v,\frac 1m} &=& -\frac{1}{48}\left( (D+3F)^2-3 (D-F)^2\right)
    \frac{1}{m_B} +\frac{1}{36}|C|^2 \frac{1}{m_B}\left(Z+\frac 52\right)
    \left(\frac 12 -Z\right).
   \ee
We shall fix $Z$ from $\pi N$ scattering ignoring $SU(3)$ breaking which
occurs at higher chiral order than we need. The precise value of $Z$
turns out not to be important for both kaonic atoms and kaon condensation.
Now our chiral Lagrangian gives, at tree order, 
the isoscalar $\pi N$ scattering length 
   \be
   a^{(+)}_{\pi N} &=& \frac{1}{4\pi f^2 (1+M_\pi /m_B)} 
    ( 2\tilde D_{\pi N} M_\pi^2 +\Sigma_{\pi N} )
   \ee
where $\Sigma_{\pi N}$ is the $\pi N$ sigma term $(\approx 45 MeV)$ and
$\tilde D_{\pi N} =\frac 14 (d_1+d_2) +\frac 12 (d_5+d_6)$. If
we take the empirical value of $a^{(+)}_{\pi N} =-0.01 M_\pi^{-1} $, we obtain
   \be
   \tilde D_{\pi N}^{emp} \approx -1.29/m_B \approx -0.27 \fm.\label{Dexp}
   \ee
The $1/m$ correction to $\pi N$ scattering is
   \be
   \tilde D^{\frac 1m}_{\pi N} = -\frac{(D+F)^2}{16} \frac{1}{m_B}
   -\frac 29 |C|^2\frac{1}{m_B}\left(Z+\frac 52\right)\left(\frac 12 -Z\right).
   \ee
For the constants $D$, $F$ and  $C$ used in this paper, this formula
reproduces the experimental value (\ref{Dexp}) for $Z\simeq -0.5$.
This will be used in our calculation of the off-shell amplitudes.

Alternatively one could fix $Z$ at one-loop order as in \cite{meissner}.
This however affects our results insignificantly. To see this, take 
$Z\approx 0.15$ obtained in ref.\cite{meissner}. The resulting off-shell
$KN$ amplitudes are given by the dotted lines in Fig.2. This represents
less than 1 \% change in the kaon self-energy, an uncertainty that can be
safely ignored in the noise of other uncertainties inherent in heavy-baryon
chiral perturbation theory.


\renewcommand{\theequation}{E.\arabic{equation}}
\setcounter{equation}{0}
\section*{ Appendix E: Off-Shell Amplitudes}
\indent

In terms of the low-energy parameters fixed by the on-shell constraints,
the off-shell $K^-N$ scattering amplitude is given by
   \beq
    a^{K^-p} &=&\left.  \frac{1}{4\pi (1+\omega/m_B)} \right\{
      T_v^{K^-p}(\omega=M_K)
     - \frac{\omega^2}{f^2}
        \left(\frac{\bar g_{{\Lambda^\star}_R}^2}{\omega+m_B-m_{{\Lambda^\star}_R}} \right)
   \nonumber\\ && \;\;\;\;\;\;\;\;\;\;
    +\frac{1}{f^2} (\omega-M_K) +\frac{1}{f^2} (\omega^2-M_K^2)
   \left(\bar d_s-\frac{\sigma_{KN}}{M_K^2} +\bar d_v +\frac{(\hat m+m_s)
   a_1}{2M_K^2}\right)
   \nonumber\\ && \;\;\;\;\;\;\;\;\;\; \left.
      +\frac{1}{f^2} ( L^+_p (\omega) -L^+_p(M_K))
    -\frac{1}{f^2} ( L^-_p (\omega) -L^-_p (M_K))  \right\},
   \label{app}\\
    a^{K^-n} &=& \left. \frac{1}{4\pi (1+\omega/m_B)} \right\{
    T_v^{K^-n}(\omega=M_K)
   \nonumber\\ && \;\;\;\;\;\;\;\;\;\;
    \frac{1}{2f^2} (\omega-M_K)
    +\frac{1}{f^2}({\omega}^2 -M_K^2)
   \left(\bar d_s -\frac{\sigma_{KN}}{M_K^2} -\bar d_v
     -\frac{(\hat m+m_s)a_1}{2M_K^2}\right)
   \nonumber\\ && \;\;\;\;\;\;\;\;\;\; \left.
     + \frac{1}{f^2}(L^+_n (\omega) -L^+_n(M_K))
    -\frac{1}{f^2} (L^-_n(\omega)-L^-_n(M_K))   \right\}.
   \label{apn}\eeq
Here
   \beq
   L^+_p(\omega)
     &=&\frac{\omega^2}{64\pi f^2} \Big\{
     \Big[2(D-F)^2+\frac 13(D+3F)^2\Big] M_K
   + \frac{3}{2} (D+F)^2 M_\pi
   \nonumber\\ &&  \;\;\;\;\; \;\;\;\;\;
   +\frac 12 (D-3F)^2 M_\eta
   -\frac {1}{3}(D+F)(D-3F)( M_\pi+M_\eta)
   \nonumber\\
   && \;\;\;\;\; \;\;\;\;\;
   -\frac {1}{6}(D+F)(D-3F) (M_\pi^2+M_\eta^2)\int^1_0 dx \frac{1}{
    \sqrt{(1-x)M_\pi^2+ xM_\eta^2}}  \Big\} \nonumber\\
   && +\frac{\omega^2}{8 f^2}
     \left(4\Sigma_K^{(+)}(-\omega)+  5\Sigma_K^{(+)}(\omega)
      +2\Sigma_\pi^{(+)}(\omega)
   +3\Sigma_\eta^{(+)}(\omega)\right),
   \nonumber\\
   L^-_p(\omega) &=& \alpha_p M_K^2 \omega +\beta_p \omega^3
    +\frac{1}{4f^2} \omega^2 \left\{-\frac 12\Sigma_K^{(-)}(\omega)
    -\Sigma_\pi^{(-)}(\omega)-\frac 32\Sigma_\eta^{(-)}(\omega)\right\},
   \nonumber\\
   L^-_p(M_K) &=& (\bar g_s+\bar g_v) M_K^3,\nonumber\\
   L^+_n(\omega)
     &=&\frac{1}{64\pi f^2}\omega^2 \Big\{
     \Big[\frac 52 (D-F)^2+\frac 16(D+3F)^2\Big] M_K
   +  \frac 32 (D+F)^2 M_\pi
    \nonumber\\
   && \;\;\;\;\; \;\;\;\;\;
   +\frac 12 (D-3F)^2 M_\eta
    +\frac 13 (D+F)(D-3F)(M_\pi+M_\eta)
   \nonumber\\ && \;\;\;\;\; \;\;\;\;\;
    +\frac 16 (D+F)(D-3F)(M_\pi^2+M_\eta^2) \int^1_0 dx \frac{1}{
    \sqrt{(1-x)M_\pi^2+x M_\eta^2}} \Big\} \nonumber\\
   &&+ \frac{\omega^2}{8 f^2} \cdot
     \left(2\Sigma_K^{(+)}(-\omega)+
      \Sigma_K^{(+)}(\omega) +\frac 52 \Sigma_\pi^{(+)}(\omega)
    +\frac 32 \Sigma_\eta^{(+)}(\omega) \right),
   \nonumber\\
   L^-_n(\omega) &=& \alpha_n M_K^2 \omega +\beta_n \omega^3
    +\frac{1}{4f^2} \omega^2 \left\{\frac 12\Sigma_K^{(-)}(\omega)
    -\frac 54\Sigma_\pi^{(-)}(\omega)-\frac 34\Sigma_\eta^{(-)}(\omega)\right\},
   \nonumber\\
   L_n^-(M_K) &=& (\bar g_s-\bar g_v) M_K^3,
   \eeq
where
   \beq
   \Sigma_i^{(+)} (\omega)
     &=& -\frac{1}{4\pi} \sqrt{M_i^2-\omega^2} \times \theta(M_i-|\omega|)
     +\frac{i}{2\pi}\sqrt{\omega^2-M_i^2}\times \theta(\omega-M_i),
   \nonumber\\
   \Sigma_i^{(-)} (\omega)&=&
     -\frac{1}{4\pi^2} \sqrt{\omega^2-M_i^2}\ln\left|\frac{
     \omega+\sqrt{\omega^2-M_i^2}}{\omega-\sqrt{\omega^2-M_i^2}}\right|
   \times   \theta (|\omega|-M_i)
   \nonumber\\
   &&  -\frac{1}{2\pi^2}\sqrt{M_i^2-\omega^2}\sin^{-1}
   \frac{\omega}{M_i}\times \theta(M_i-|\omega|).
\label{xxxx}   \eeq
Note that $\Sigma^\pm_i(\omega)$ result from diagrams of Fig. 1e (See
Appendix A), where intermediate
states can be real particle and result in imaginary amplitude.
The functions $L_{p,n}^-(\omega)$ contain four parameters
$\alpha_{p,n}$ and $\beta_{p,n}$. Owing to the constraints at $\omega=M_K$,
$L^-_{p,n} (M_K)$,
they reduce to two. These two cannot be fixed by on-shell data.
However since the off-shell amplitudes are
rather insensitive to the precise values of these constants, we will somewhat
arbitrarily set $\alpha_{p,n} \approx \beta_{p,n}$ in calculating Figure 2.


\renewcommand{\theequation}{F.\arabic{equation}}
\setcounter{equation}{0}
\section*{Appendix F: Interpolating Fields} 
\indent

In this Appendix, we show explicitly that to the chiral order we are concerned
with, physics does not depend on the way the kaon field $K$ (or in general
the Goldstone boson field $\pi$) interpolates. There is nothing new in what
we do below: It is a well-known theorem \footnote{A general discussion closely
related to this issue is given in a recent paper by Leutwyler \cite{leut94}.}.
But to those who are not very 
familiar with the modern notion of effective field theories, it has been a bit 
of a mystery that an off-shell amplitude which does not obey Adler's soft-pion
theorem such as in the case of our Lagrangian could give the same physics with
the amplitude which does in a situation which involves the equation of state,
not just an on-shell S-matrix. To have a simple idea, we start with the toy
model of Manohar \footnote{We wish to thank A. Manohar who showed us how to
understand the problem using this toy model.}.
\vskip 5mm

\noindent $\bullet$ {\bf Adler's Conditions in a Toy Model}
\vskip 5mm
Consider the simple Lagrangian,
   \be
   {\cal L} &=& \frac 12 \left(\partial K\right)^2
    -\frac 12 M_K^2 (1+\epsilon \bar B B )K^2
   \label{lag}\ee
where $K$ is the charged kaon field which will develop a condensate in
the form $\langle K^- \rangle  =v_K e^{-i\mu t}$.
As written,  this Lagrangian can give only a linear density
dependence in the kaon self-energy at tree order.
The effective energy-density for s-wave kaon condensation linear in
$v_K^2$ is
   \be
   \tilde \epsilon 
   &=& ( \mu^2 -M_K^2 (1+\epsilon \rho_N) ) v_K^2.
   \label{ee1}\ee
Thus the chemical potential in the ground state is
   \be
   \mu^2 =M_K^2 (1+\epsilon \rho_N).
   \ee
We shall now redefine the kaon field. To do this, we have, from the
Noether construction and the equation of motion,
   \be
   j_A^\mu &=& f \partial^\mu K, \nonumber\\
   \partial_\mu j^\mu_A &=& f M_K^2 (1+\epsilon\bar B B) K .
   \ee
This invites us to redefine the kaon field as
   \be
   \tilde K = K (1+\epsilon \bar B B ) = 
     \frac{\partial_\mu j_A^\mu}{f M_K^2}
   \label{defK}\ee
thus giving the divergence of the axial current as the interpolating field
of the kaon field. This kaon field will then satisfy the Adler soft-meson
theorem. To see this, we rewrite the original Lagrangian in terms of 
the new field $\tilde K$, 
   \be
   {\cal L} &=&\frac{1}{(1+\epsilon \bar BB)^2}\left(
    \frac 12 \left(\partial \tilde K\right)^2
    -\frac 12 M_K^2 (1+\epsilon \bar B B )\tilde K^2 \right)
    +{\cal O} (\partial B)
   \label{lag2} \\
    &=&
    \frac 12 {(1-2\epsilon \bar BB)}\left(\partial \tilde K\right)^2
    -\frac 12 M_K^2 (1-\epsilon \bar B B )\tilde K^2 
   +{\cal O}\left(\tilde K^2  (\bar B B)^{n\ge 2} \right)
    +{\cal O} (\partial B).
   \nonumber \ee
We can immediately read off the $KN$ scattering amplitude and verify that
it satisfies the Adler theorem. Note, however, that to obtain the same
kaon self-energy that enters into the energy-density of the matter
system as the one given by eq.(\ref{lag}), it would be necessary to
keep multi-baryon terms, which means that in medium, higher density dependence
needs to be taken into account. 

Now to see that the interpolating fields $K$ and $\bar{K}$ give the same physics
in medium, consider the critical point at which condensation sets in.
{}From eq.(\ref{defK}), we have 
   \be
   |\langle \tilde K \rangle| &=& \tilde v_K = v_K ( 1+\epsilon \rho_N).
   \ee
Substituting this into eq.(\ref{lag}) or directly from eq.(\ref{lag2}),
we get the effective energy-density 
   \be
   \tilde \epsilon &=& (\mu^2 -M_K^2 (1+\epsilon\rho_N))
   \frac{\tilde v_K^2}{(1+\epsilon \rho_N)^2}.
   \label{ee2}\ee
Clearly (\ref{ee1}) and (\ref{ee2}) give the same critical point given by
the vanishing of the energy density. This is of course quite trivial.
If one keeps terms up to linear in $\rho_N$, eq.(\ref{ee2}) reads
   \be
   \tilde \epsilon &=& 
   (\mu^2 -M_K^2 -(2\mu^2 -M_K^2)\epsilon \rho_N ) \tilde v_K^2.
   \label{ee3}\ee
This is the kaon self-energy in linear density approximation, which satisfies
Adler's consistency condition. As it stands, it looks very different
from eq.(\ref{ee1}), but to the linear order in density,
this gives the same pole position,
   \be
   (\mu^2 -M_K^2 -(2\mu^2 -M_K^2)\epsilon \rho_N ) = 0 
   & \longrightarrow &
   (1-2\epsilon\rho_N) \mu^2 = M_K^2 (1-\epsilon\rho_N) 
   \nonumber\\
   & \longrightarrow &
   \mu^2  = M_K^2 (1+\epsilon\rho_N).
   \ee
Beyond the linear-density approximation, one must use eq.(\ref{ee2})
instead of eq.(\ref{ee3}) for the physics of the toy-model Lagrangian: 
Consistency requires that
{\it all} kaon-multi-nucleon scattering terms 
(that is, terms higher order in $\rho_N$) be included.

We now turn to the real issue. We shall compare the approach used here
(called ``Kaplan-Nelson" (KN)) to the
Gasser-Sainio-Svarc (GSS) approach which implements Adler's conditions by means
of external fields \cite{GSS}.
\vskip 5mm
\noindent $\bullet$ {\bf The KN Approach }
\vskip 5mm

The KN approach corresponds to introducing the source field $p_i$
in the Lagrangian
   \be
   {\cal L}_{source} &=& p^+ K^-  + p^- K^+ \cdots.
   \ee
The self-energy is then obtained from
   \be
   \langle \tau(x_1,x_2)\rangle  &=& \langle T ( K^+ (x_2) K^- (x_1))
    \rangle_\rho \nonumber\\
   &=& -\left\langle
   \left. \frac{\delta^2}{\delta p^+(x_1) \delta p^-(x_2)} \int {\cal D}[K,B] 
   e^{i \int ({\cal L}_{orig} +{\cal L}_{source} 
   ) }\right|_{p^{\pm}=0,\cdots=0} \right\rangle_\rho
   \nonumber\\ 
   &=& i\Delta^{(0)}_K(x_1-x_2) +i \Sigma_K \int \Delta^{(0)}_K(x_1 -z) 
\Delta^{(0)}_K(z-x_2) + {\cal O} (\Sigma_K^2) 
   \label{two}\ee
where  $\langle \cdots \rangle_\rho $ represents the expectation value 
in the ground state of dense matter. Here $\Delta^{(0)}_K$ is the free kaon 
propagator 
   \be
   \Delta^{(0)}_K(x-y) =\frac{1}{(2\pi)^4}\int \frac{e^{-ik\cdot (x-y)}}{
    k^2-M_K^2 +i\epsilon} d^4 k
   \ee
and $\Sigma_K$ is the kaon self-energy calculated to order $\nu=3$
with the Lagrangian
   \be {\cal L}_0 &=& \partial K^+\partial K^- -M_K^2 K^+ K^- \nonumber\\
    {\cal L}_{int} &=& {\cal L}_{int}^{\nu=1} +{\cal L}_{int}^{\nu=2} + \cdots.
   \label{pert}\ee
Summing the series (\ref{two}), we have
   \be
   \tau (x_1,x_2)
   &=& \frac{i}{(2\pi)^4} \int \frac{e^{ik\cdot (x_1-x_2)}}{k^2-M_K^2-
   \Sigma_K +i\epsilon} d^4 k.
   \ee
The propagator to order $\nu=3$ in momentum space is
   \be
    \Delta_K &=& \frac{i}{k^2-M_K^2-\Sigma_K+i\epsilon}
   \label{pro1}\ee
and the effective energy density to order $v_K^2$ is
   \be
   \epsilon &=& - i  \Delta_K^{-1} |v_K|^2 \nonumber\\
    &=& -( \mu^2-M_K^2 -\Sigma_K ) |v_K|^2.
   \ee

\vskip 5mm
\noindent $\bullet$ {\bf The GSS Approach }
\vskip 5mm

The GSS approach corresponds to introducing into the Lagrangian 
the source field $p_i$
of the form
   \be
   {\cal L}_{source} &=& p^+ K^- (1+\alpha_p\bar p p+\alpha_n \bar n n) 
   + p^- K^+ (1+\alpha_p\bar p p+\alpha_n \bar n n) + \cdots
   \ee
where $ \alpha_p = \frac{1}{f^2 M_K^2} (\sigma_{KN} +C_{KN})$ and
$ \alpha_n = \frac{1}{f^2 M_K^2} (\sigma_{KN} -C_{KN})$ are related with 
the parameters of the Lagrangian (\ref{L2}) by
 $$\sigma_{KN}=-\frac 12 (\hat m+m_s) (a_1+2a_2+4a_3)$$
and $$C_{KN}=-\frac 12 (\hat m+m_s) a_1.$$
As shown in ref.\cite{GSS}, this way of introducing the source field
reproduces Adler's conditions (soft-pion, self-consistency etc.).
Now the self-energy is obtained from the two-point function
   \be
   \langle \tau(x_1,x_2)\rangle 
   &=&
   -\left\langle
   \left. \frac{\delta^2}{\delta p^+(x_1) \delta p^-(x_2)} \int {\cal D}[K,B] 
   e^{i \int ({\cal L}_{orig} +{\cal L}_{source} ) }\right
   |_{p^{\pm}=0,\cdots=0} \right\rangle_\rho
   \nonumber\\
   &=& 
   \langle T ( K^+ (1+\alpha_p\bar p p+\alpha_n\bar n n) (x_2)
   K^- (1+\alpha_p\bar p p+\alpha_n\bar n n) (x_1) \rangle_\rho
   \nonumber\\
   &=& (1+\alpha_p \rho_p+\alpha_n \rho_n)^2 
   \langle  T ( K^+(x_2) K^- (x_1)) \rangle_\rho
   \nonumber\\
   &=& (1+\alpha_p \rho_p+\alpha_n \rho_n)^2 
    \frac{i}{(2\pi)^4} \int \frac{e^{ik\cdot (x_1-x_2)}}{k^2-M_K^2-
   \Sigma_K +i\epsilon} d^4 k
   \ee
where $\Sigma_K$ is identical to that of the KN approach since the $\alpha_p$
and $\alpha_n$ are $\nu=2$ terms that are unaffected by loop corrections (of
order $\nu=3$).
Finally the full propagator is given by
   \be
    \Delta_K^\prime &=& (1+\alpha_p\rho_p +\alpha_n\rho_n)^2 
   \frac{i}{k^2-M_K^2-\Sigma_K+i\epsilon}
   \label{pro}\ee
which can be rewritten in the form of eq.(\ref{pro1});
   \be
    \Delta_K^\prime &=&  \frac{i}{k^2-M_K^2-\Sigma^\prime_K+i\epsilon}
   \ee
with a redefined self-energy 
   \be
   \Sigma^\prime_K &=& \frac{1}{(1+\alpha_p\rho_p +\alpha_n\rho_n)^2}\left\{
   (k^2-M_K^2)\left[(1+\alpha_p\rho_p +\alpha_n\rho_n)^2 -1
	       \right] + \Sigma_K \right\} .
   \ee
One can verify that
to the linear order in density, this has the structure mentioned in the
main text, namely, it changes from an attraction on-shell to a repulsion
off-shell.
Finally the effective energy density linear in $\tilde v_{K}^2$ is 
   \be
   \tilde \epsilon &=& - i ({\Delta_K^\prime})^{-1} |\tilde v_{K}|^2
    = -i \Delta_K^{-1} \left|\frac{\tilde v_{K}}
   {1+\alpha_p\rho_p +\alpha_n\rho_n}\right|^2 
   \ee
where $\tilde v_K$ is the kaon expectation value in terms of the GSS field.

\vskip 5mm
\noindent $\bullet$ {\bf Effective energy-density and physical observables}
\vskip 5mm

As in the toy model, we can write $\tilde v_{ K}$ in terms of $v_K$:
   \be
   \tilde v_{K} &=& v_K ( 1 + a x +b)
   \label{transf}\ee
where $a=(\alpha_p-\alpha_n)\rho$ and $b=\alpha_n \rho$.
Thus the energy-densities to all orders in $v_K$ and $\tilde v_{K}$ 
can be written as
   \be
   \epsilon &=& \sum_{n\ge 2} \alpha_n (\mu,x,\rho) |v_K|^n +\beta(\mu,x,\rho), 
   \label{c1}\\
   \tilde \epsilon &=& \sum_{n\ge 2} \alpha_n (\mu,x,\rho) 
    \left|\frac{\tilde v_K}{1+ a x+b} \right|^n +\beta(\mu,x,\rho) 
   \label{c2}\ee
where $a_n(\mu,x,\rho)$ and  $\beta(\mu,x,\rho)$ are given in eq.(\ref{effen}).
By differentiating the effective energy density with respect to
$\mu,x,v_K$ $(\tilde v_K)$, 
one can get the equation of state as a function of density $\rho$.
In the GSS approach, we have three equations of state;
   \be
   0=\frac{\partial\tilde \epsilon}{\partial\mu} &=& \sum_{n\ge 2} 
   \frac{\partial \alpha_n(\mu,x,\rho)}{\partial \mu} 
   \left(\frac{\tilde v_K}{1+ax+b}\right)^n
   +\frac{\partial \beta(\mu,x,\rho)}{\partial \mu} 
   \nonumber\\
   0=\frac{\partial\tilde \epsilon}{\partial x} &=& \sum_{n\ge 2} 
   \frac{\partial \alpha_n(\mu,x,\rho)}{\partial  x} 
   \left(\frac{\tilde v_K}{1+ax+b}\right)^n
   +\frac{\partial \beta(\mu,x,\rho)}{\partial  x}
   \nonumber\\ &&
   + \frac{a}{1+ax+b} \sum_{n\ge 2} n  \alpha_n(\mu,x,\rho)
   \left(\frac{\tilde v_K}{1+ax+b}\right)^n 
   \nonumber\\
   0=\frac{\partial \epsilon}{\partial \tilde v_K} &=& 
   \frac{1}{\tilde v_K} \sum_{n\ge 2} n \alpha_n(\mu,x,\rho)
   \left(\frac{\tilde v_K}{1+ax+b}\right)^n .
   \label{ca2}\ee
Now using the third equation, we see that the last term in the second equation
vanishes.  To see that the two ways give the same physics, it suffices to note 
that the KN approach gives the same set of equations
except for the replacement $\tilde v_{ K} = v_K ( 1+ ax +b)$ which 
does not affect anything.

An identical conclusion is reached by Thorsson and Wirzba \cite{thorsson} 
by a slightly different reasoning.


\renewcommand{\theequation}{G.\arabic{equation}}
\setcounter{equation}{0}
\section*{Appendix G: In-Medium Modifications}
\indent

In this Appendix we write down the explicit forms of the quantities that
enter in eq.(\ref{self2}). 
    \be
    \delta {\cal T}^{K^-p}_{\rho_N}
    &=& -\left. \frac{1}{12f^4}\right\{ (D+F)^2 \left[(M_\pi^2+M_K^2-\omega^2)
      D_{4,\pi\pi}^n-D_{6,\pi\pi}^n\right]
    \nonumber\\
    &&
    +\frac 12 (D+F)^2 \left[(M_\pi^2+M_K^2-\omega^2) D_{4,\pi\pi}^p
    -D_{6,\pi\pi}^p
    \right]
    \nonumber\\
    &&  +\frac 12 (D-3 F)^2\left[\left(M_K^2-\frac 13 M_\pi^2-\omega^2\right)
    D_{4,\eta\eta}^p-D_{6,\eta\eta}^p \right]
    \nonumber\\
    && \left. -(D+F) (D-3F)\left[
      \left( \frac 13 M_\pi^2-\frac 13 M_K^2-\omega^2\right)
     D_{4,\pi\eta}^p -D_{6,\pi\eta}^p \right] \right\}
    \nonumber\\
    && +  \frac{\omega^2}{4f^4}\left( 8 \Sigma_{K}^p(\omega)
          + \Sigma_{K}^n (\omega) \right)
     + \frac{1}{12 f^4}\left\{ 2(D+F) F M_\pi^2\Sigma_\pi^{p}(0) \right.
    \nonumber\\
    && \left.
     +(D+F)^2 M_\pi^2 \Sigma_\pi^{n}(0)
   -2(D-3F)F M_\eta^2 \Sigma_\eta^{p}(0) \right\}
   \\
   \delta{\cal T}^{K^-n}_{\rho_N}
   &=& -\left. \frac{1}{12f^4}\right\{ (D+F)^2 \left[(M_\pi^2+M_K^2-\omega^2)
      D_{4,\pi\pi}^p-D_{6,\pi\pi}^p\right]
   \nonumber\\
   &&
   +\frac 12 (D+F)^2 \left[(M_\pi^2+M_K^2-\omega^2) D_{4,\pi\pi}^n
   -D_{6,\pi\pi}^n
   \right]
   \nonumber\\
   &&  +\frac 12 (D-3 F)^2\left[\left(M_K^2-\frac 13 M_\pi^2-\omega^2\right)
     D_{4,\eta\eta}^n-D_{6,\eta\eta}^n \right]
     \nonumber\\
     && \left. +(D+F) (D-3F)\left[
   \left( \frac 13 M_\pi^2-\frac 13 M_K^2-\omega^2\right)
     D_{4,\pi\eta}^n -D_{6,\pi\eta}^n \right] \right\}
    \nonumber \\
     && +  \frac{\omega^2}{4f^4}\left( 2 \Sigma_{K}^{n} (\omega)
            + \Sigma_{K}^{p} (\omega) \right)
    + \frac{1}{12 f^4}\left\{ (D^2-F^2) M_\pi^2\Sigma_\pi^{n}(0) \right.
   \nonumber\\ && \left.
   +(D+F)^2 M_\pi^2 \Sigma_\pi^{p}(0)
    +(D-3F)(D-F) M_\eta^2 \Sigma_\eta^{n}(0) \right\},
   \\
    D_{\alpha,ij}^N &=& \frac{1}{2\pi^2}\int_0^{k_{F_N}} d|\vec k|
      \frac{1}{|\vec k|^2+M_i^2}\frac{1}{|\vec k|^2+M_j^2}
      |\vec k|^\alpha
    \nonumber\\
    \Sigma_i^{N}(\omega) &=&
      \frac{1}{2\pi^2} \int_0^{k_{F_N}} d|\vec k|
      \frac{|\vec k|^2}{\omega^2-M_i^2-|\vec k|^2}.
   \ee
Here the subscripts $\pi$, $\eta$ and $K$ are the octet Goldstone bosons, the
superscripts $n$ and $p$ stand for neutrons and protons.


\newpage


  \begin{center}
  Table 1 : \parbox[t]{4in}{  
       Scattering lengths from three leading order contributions
       for the empirical value of the constant $g_{\Lambda^\star}^2 =0.25$.
        Also shown is the contribution from $\Lambda^\star$.}
  \end{center}
  $$
  \begin{array}{|r||r|r|r|r|}
  \hline
  {g^2_{\Lambda^\star}=0.25} &{\cal  O}(Q) & {\cal O}(Q^2) & {\cal O}(Q^3) &
  \Lambda^\star \\
  \hline
  \hline
  a^{K^+p} (fm) & -0.588 & 0.316 & -0.114 & 0.076 \\ \hline
  a^{K^-p} (fm) &0.588 & 0.316 & -0.143 & -1.431 \\ \hline
  a^{K^+n} (fm) & -0.294 & 0.277 & -0.183 & 0.000\\ \hline
  a^{K^-n} (fm) & 0.294 & 0.277 & -0.201 & 0.000\\ \hline
  \end{array}
  $$


\vskip 2cm
     \begin{center}
     Table 2 :\parbox[t]{5.5in}{ 
     Self energies for kaonic atoms in nuclear matter ($x=0.5$) in unit of
     $M_K^2$ at $u=0.97$ for $g_{\Lambda^\star}^2 =0.25$. 
     $\Delta V=M_K^*-M_K$ is the attraction (in unit of MeV) at a given 
     $(C_{\Lambda^\star}^S-C_{\Lambda^\star}^T) f^2$.}
     \end{center}
    $$
    \begin{array}{|c||c|c||c|c|c|c|}
    \hline
    ( C_{\Lambda^\star}^S-C_{\Lambda^\star}^T) f^2 & M_K^* & \Delta V &  
    -\rho{\cal T}^{free} & -\rho\delta {\cal T}^{free}
    & \Pi_{\Lambda^\star}^1 &\Pi_{\Lambda^\star}^2 \\ \hline \hline
       1 &  396.4& -98.65&  -0.2361& 0.04485& 
       -0.2113& 0.04344 \\ \hline
       5 &  351.4& -143.6&  -0.1810& 0.03111& 
       -0.3541& 0.00903 \\ \hline
      10 &  326.2& -168.8&  -0.1738& 0.02637& 
      -0.4215& 0.00418 \\ \hline
      50 &  253.9& -241.1&  -0.1842& 0.01839& 
      -0.5715& 0.00056 \\ \hline
     100 &  218.7& -276.3&  -0.1927& 0.01621& 
     -0.6281& 0.00021 \\ \hline
     \end{array}
     $$


\vskip 2cm
    \begin{center}
    Table 3 :\parbox[t]{5.0in}{
     Self-energies for kaonic atoms in nuclear matter ($x=0.5$)
    in unit of $M_K^2$ for $g_{\Lambda^\star}^2=0.25$ and
    $(C_{\Lambda^\star}^S-C_{\Lambda^\star}^T) f^2 =10$.
    $\Delta V\equiv M_K^\star -M_K$ is the attraction (in unit of MeV) at
    given density.}
    \end{center}
    $$
    \begin{array}{|c||c|c||c|c|c|c|}
    \hline
     u & M_K^* & \Delta V & -\rho {\cal T}^{free} & -\rho \delta {\cal T}^{free}
    & 
    \Pi^1_{\Lambda^\star} & \Pi^2_{\Lambda^\star} \\
    \hline
    \hline
     0.2& 424.6& -70.37&  -0.0673& 0.0034& -0.1998& 0.007607\\
     0.4& 390.0& -105.0&  -0.0920& 0.0084& -0.3024& 0.006641\\
     0.6& 364.3& -130.7&  -0.1173& 0.0143& -0.3610& 0.005635\\
     0.8& 342.6& -152.4&  -0.1462& 0.0207& -0.3996& 0.004794\\
     1.0& 323.5& -171.5&  -0.1789& 0.0274& -0.4250& 0.004088\\
     1.2& 306.2& -188.8&  -0.2150& 0.0343& -0.4404& 0.003483\\
     1.4& 289.9& -205.1&  -0.2540& 0.0413& -0.4460& 0.002945\\
    \hline
    \end{array}
    $$

\newpage

\centerline{Table 4 :  Numerical values in MeV of $r$ as 
function of $x$ and $u$.}
$$
\begin{array}{|c|c|c|c|} \hline
 r(u,x)  & x=0.0 & x= 0.5 & x=1.0 \\ \hline
u=  0.5 & 30.78 & 38.25 & 30.78 \\ \hline
 u= 1.0 & 47.52 & 61.55 & 47.52 \\ \hline
u=  1.5 & 60.20 & 79.75 & 60.20 \\ \hline
\end{array}
$$


\vskip 2cm
\begin{center}
Table 5 : \parbox[t]{5.0in}{
  Self-energies for $K^+$ in nuclear matter ($x=0.5$)
in unit of $M_K^2$ for $g_{\Lambda^\star}^2=0.25$ and
$(C_{\Lambda^\star}^S-C_{\Lambda^\star}^T) f^2 =10$.
$\Delta V\equiv M_K^\star -M_K$ is the repulsion (in unit of MeV) at
given density.}
\end{center}
$$
\begin{array}{|c||c|c||c|c|c|c|}
\hline
 u & M_K^* & \Delta V &  -\rho {\cal T}^{free} & -\rho \delta {\cal T}^{free}
& 
\Pi^1_{\Lambda^\star} & \Pi^2_{\Lambda^\star} \\
\hline
\hline
 0.2& 507& 12& 0.025& 0.019& -0.0009& 0.0002 \\ 
 0.4& 515& 20& 0.048& 0.040& -0.0038& 0.0005 \\ 
 0.6& 523& 28& 0.068& 0.053& -0.0087& 0.0006 \\ 
 0.8& 532& 37& 0.085& 0.054& -0.0157& 0.0006 \\ 
 1.0& 537& 42& 0.102& 0.083& -0.0248& 0.0010 \\ 
 1.2& 542& 47& 0.117& 0.112& -0.0360& 0.0014 \\
 1.4& 547& 52& 0.129& 0.122& -0.0494& 0.0015 \\
\hline
\end{array}
$$


\vskip 2cm
   \begin{center}
   Table 6 : \parbox[t]{4in}{ Critical density $u_c$ in 
   in-medium two-loop chiral
   perturbation theory for $g_{\Lambda^\star}^2=0.25$. (a) correspond to
   linear density approximation of Fig .3a and (b) to the full two-loop result
   for $(C_{\Lambda^\star}^S-C_{\Lambda^\star}^T) f^2=10$.}
   \end{center}
    $$
    \begin{array}{|c||c||c|c|c|}
    \hline
     & & \multicolumn{3}{c|}{(b)} \\ \cline{3-5}
     \hphantom{a} F(u) \hphantom{a} & \hphantom{aaa} (a) \hphantom{aaa}  
     &C_{\Lambda^\star}^S f^2 =10 & C_{\Lambda^\star}^S f^2 =5 
     & C_{\Lambda^\star}^S f^2 =0 \\
    \hline
    \hline
    \frac{2 u^2}{1+u} & 3.77 & 2.81 & 2.98 & 3.24 \\ \hline
          u        & 3.90 & 3.13 & 3.33 & 3.69 \\ \hline
       \sqrt{u}    & 4.11 & 3.71 & 3.96 & 4.41 \\ \hline
    \end{array}
    $$

\newpage

   \begin{center}
    Table 7 :  \parbox[t]{4in}{ Critical density $u_c$ in 
   in-medium two-loop chiral
   perturbation theory for $g_{\Lambda^\star}^2=0.25$ and $F(u)=u$.}
   \end{center}
   $$\begin{array}{|c||c|c|c|}
   \hline
  (C_{\Lambda^\star}^S -C_{\Lambda^\star}^T) f^2  
   & C_{\Lambda^\star}^S f^2 =100 
   &C_{\Lambda^\star}^S f^2 =10 &C_{\Lambda^\star}^S f^2 =0 \\
   \hline \hline
   1 & 2.25  & 3.25 & 4.51 \\ \hline
   10 & 2.24  & 3.13 & 3.69 \\ \hline
   100 & 2.18  & 2.66 & 2.77 \\ \hline
   \end{array}
   $$


\vskip 2cm
    \begin{center}
    Table 8 :\parbox[t]{5.5in}{ Self-energies for kaonic atoms in 
    nuclear matter ($x=0.5$) in unit of
    $M_K^2$ at $u=0.97$ for $g_{\Lambda^\star}^2 =0.05$. 
    $\Delta V=M_K^*-M_K$ is the attraction (in unit of MeV) 
    at a given $(C_{\Lambda^\star}^S-C_{\Lambda^\star}^T) f^2$.}
    \end{center}
   $$
   \begin{array}{|c||c|c||c|c|c|c|}
   \hline
   ( C_{\Lambda^\star}^S-C_{\Lambda^\star}^T) f^2 & M_K^* & \Delta V &  
   -\rho{\cal T}^{free} & -\rho\delta {\cal T}^{free}
   & \Pi_{\Lambda^\star}^1 &\Pi_{\Lambda^\star}^2 \\
   \hline
   \hline
     10&  386& -109&  -0.114& 0.0407&  -0.3187& 0.001164\\ \hline
     20&  364& -131&  -0.116& 0.0342&  -0.3773& 0.000548\\ \hline
     50&  330& -165&  -0.128& 0.0270&  -0.4535& 0.000187\\ \hline
     70&  316& -179&  -0.135& 0.0248&  -0.4822& 0.000123\\ \hline
    100&  300& -195&  -0.143& 0.0227&  -0.5129& 0.000079\\ \hline
    \end{array}
    $$


\vskip 2cm
   \begin{center}
   Table 9 : \parbox[t]{3.5in}{ Critical density $u_c$ 
   in in-medium two-loop ChPT
   for $(C_{\Lambda^\star}^S-C_{\Lambda^\star}^T) f^2 =70$, 
   $g_{\Lambda^\star}^2=0.05$.}
   \end{center}
   $$
   \begin{array}{|c||c|c|c|c|} \hline
   \hphantom{a} F(u)\hphantom{a}  &   C_{\Lambda^\star}^S f^2 =70 
   & C_{\Lambda^\star}^S f^2 =40 & C_{\Lambda^\star}^S f^2 =10 
    & C_{\Lambda^\star}^S f^2 =0 \\ 
    \hline
    \hline
   \frac{2 u^2}{1+u} &  2.68 & 2.83 & 3.04 & 3.14 \\ \hline
    u                &  2.99 & 3.17 & 3.45 & 3.59 \\ \hline
    \sqrt u          &  3.56 & 3.79 & 4.15 & 4.35 \\ \hline
   \end{array}
   $$

\newpage

   \begin{center}
   Table 10 : \parbox[t]{3.5in}{Critical density $u_c$ in in-medium 
   two-loop ChPT for $g_{\Lambda^\star}^2=0.05$  and $F(u)=u$.}
   \end{center}
   $$
   \begin{array}{|c||c|c|c|} \hline
    ( C_{\Lambda^\star}-C_{\Lambda^\star}^T) f^2 & C_{\Lambda^\star}^S f^2 = 100 
     & C_{\Lambda^\star}^S f^2 = 10 & C_{\Lambda^\star}^S f^2 = 0 \\
   \hline \hline
   1 &  2.95 & 4.07 & 5.34(?) \\ \hline
   10 &  2.94 & 3.91 & 4.44 \\ \hline
   100 & 2.83 & 3.33 & 3.44 \\ \hline
   \end{array}
   $$


\vskip 2cm
     \begin{center}
     Table 11 :\parbox[t]{5.5in}{ 
     Self-energies for kaonic atoms {\it with BR scaling}
      in nuclear matter ($x=0.5$) in unit of
     $M_K^2$ at $u=0.97$ for $g_{\Lambda^\star}^2 =0.25$. $\Delta V=M_K^*-M_K$ 
     is the attraction (in unit of MeV) at a given 
     $(C_{\Lambda^\star}^S-C_{\Lambda^\star}^T) f^2$.}
     \end{center}
    $$
    \begin{array}{|c||c|c||c|c|c|c|}
    \hline
    ( C_{\Lambda^\star}^S-C_{\Lambda^\star}^T) f^2 & M_K^* & \Delta V &  
    -\rho{\cal T}^{free} & -\rho\delta {\cal T}^{free}
    & \Pi_{\Lambda^\star}^1 &\Pi_{\Lambda^\star}^2 \\
    \hline
    \hline
  1&  386.5& -108.5&  -0.2985& 0.04102&  -0.1626& 0.02995\\ \hline
  5&  344.5& -150.5&  -0.2596& 0.02966&  -0.3056& 0.00727\\ \hline
 10&  319.5& -175.5&  -0.2550& 0.02533&  -0.3698& 0.00343\\ \hline
 50&  246.3& -248.7&  -0.2677& 0.01786&  -0.5026& 0.00045\\ \hline
100&  211.7& -283.3&  -0.2759& 0.01585&  -0.5565& 0.00017\\ \hline
     \end{array}
     $$


\vskip 2cm
    \begin{center}
    Table 12 :\parbox[t]{5.0in}{
     Self-energies for kaonic atoms {\it with BR scaling}
     in nuclear matter ($x=0.5$)
    in unit of $M_K^2$ for $g_{\Lambda^\star}^2=0.25$ and
    $(C_{\Lambda^\star}^S-C_{\Lambda^\star}^T) f^2 =10$.
    $\Delta V\equiv M_K^\star -M_K$ is the attraction (in unit of MeV) at
    given density.}
    \end{center}
    $$
    \begin{array}{|c||c|c||c|c|c|c|}
    \hline
     u & M_K^* & \Delta V & -\rho {\cal T}^{free} & -\rho \delta {\cal T}^{free}
    & 
    \Pi^1_{\Lambda^\star} & \Pi^2_{\Lambda^\star} \\
    \hline
    \hline
 0.2& 424.2& -70.75& -0.0698& 0.00340& -0.1977& 0.00749 \\ \hline
 0.4& 389.0& -106.0& -0.1032& 0.00834& -0.2946& 0.00639 \\ \hline
 0.6& 361.9& -133.1& -0.1445& 0.01405& -0.3421& 0.00521 \\ \hline
 0.8& 338.3& -156.7& -0.1982& 0.02011& -0.3655& 0.00420 \\ \hline
 1.0& 316.2& -178.8& -0.2661& 0.02625& -0.3691& 0.00331 \\ \hline
 1.2& 294.7& -200.3& -0.3498& 0.03228& -0.3569& 0.00253 \\ \hline
 1.4& 272.7& -222.3& -0.4511& 0.03798&  -0.3291& 0.00184 \\ \hline
    \end{array}
    $$

\newpage

   \begin{center}
   Table 13 : \parbox[t]{4in}{ Critical density $u_c$ {\it with BR scaling} in 
   in-medium two-loop chiral
   perturbation theory for $g_{\Lambda^\star}^2=0.25$. (a) correspond to
   linear density approximation of Fig .3a and (b) to the full two-loop result
   for $(C_{\Lambda^\star}^S-C_{\Lambda^\star}^T) f^2=10$.}
   \end{center}
    $$
    \begin{array}{|c||c||c|c|c|}
    \hline
     & & \multicolumn{3}{c|}{(b)} \\ \cline{3-5}
     \hphantom{a} F(u) \hphantom{a} & \hphantom{aaa} (a) \hphantom{aaa}  
     &C_{\Lambda^\star}^S f^2 =10 & C_{\Lambda^\star}^S f^2 =5 & C_{\Lambda^\star}^S f^2 =0 \\
    \hline
    \hline
    \frac{2 u^2}{1+u} & 2.11 & 2.08 & 2.11 & 2.14 \\ \hline
          u        & 2.16 & 2.15 & 2.17 & 2.20  \\ \hline
       \sqrt{u}    & 2.22 & 2.23 & 2.25 & 2.26 \\ \hline
    \end{array}
    $$


\vskip 2cm
   \begin{center}
    Table 14 :  \parbox[t]{4in}{ Critical density $u_c$ {\it with BR
    scaling} in in-medium two-loop chiral
   perturbation theory for $g_{\Lambda^\star}^2=0.25$ and $F(u)=u$.}
   \end{center}
   $$\begin{array}{|c||c|c|c|}
   \hline
 (C_{\Lambda^\star}^S -C_{\Lambda^\star}^T) f^2  & C_{\Lambda^\star}^S f^2 =100 
   &C_{\Lambda^\star}^S f^2 =10 &C_{\Lambda^\star}^S f^2 =0 \\
   \hline \hline
   1 & 1.92 & 2.16 & 2.21 \\ \hline
   10 & 1.92  & 2.15  & 2.20 \\ \hline
   100 & 1.89 & 2.09 & 2.12 \\ \hline
   \end{array}
   $$


\vskip 2cm
\begin{center}
Table 15 : \parbox[t]{4in}{Critical density $u_c$ {\it with BR scaling}
and without $\O (Q^3)$ terms for (A) 
$C^S_{\Lambda^\star}=C^T_{\Lambda^\star}=0$ and
(B) $(C^S_{\Lambda^\star}-C^T_{\Lambda^\star})f^2=10$. 
This should be compared with
the results of Table 13. }\end{center}
$$
\begin{array}{|c||c||c|c|c|}
\hline
     &      & \multicolumn{3}{c|}{(B)} \\
     \cline{3-5}
F(u) & \hphantom{aaa} (A)\hphantom{aaa}  & C_{\Lambda^\star}^S=100 &
C_{\Lambda^\star}^S=10 & C_{\Lambda^\star}^S=0 \\
\hline
\frac{2u^2}{1+u} &  1.89 & 1.68  & 1.85  & 1.88  \\
\hline
u & 1.94(1.944)  & 1.76  & 1.91  & 1.94(1.938)  \\
\hline
\sqrt u & 2.00(1.999)  & 1.87 & 1.98  & 2.00(1.997)  \\
\hline
\end{array}
$$

\newpage

\centerline{\bf Figure Captions}
\begin{itemize}
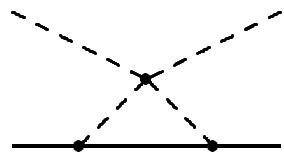
\item {\bf Figure 1:} One-loop Feynman diagrams contributing to $K^\pm N$
     scattering: The solid line represents baryons (nucleon for the external and
     octet and decuplet baryons for the internal line) and the broken line
     pseudo-Goldstone bosons ($K^\pm$ for the external and $K$, $\pi$ and
     $\eta$ for the internal line). 
     There are in total thirteen diagrams at one loop, but for
     s-wave $KN$ scattering, for reasons described in the text, we are left with
     only six topologically distinct one-loop diagrams.
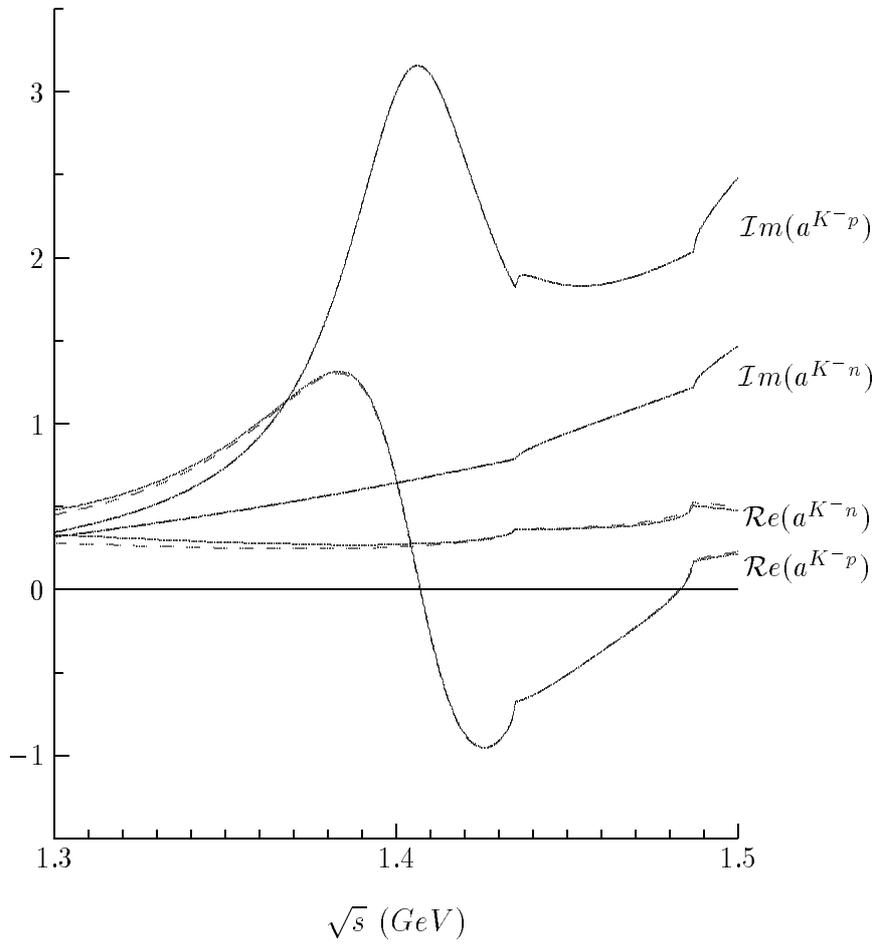
\item {\bf Figure 2:} $K^-N$ amplitudes as function of $\sqrt{s}$: These
    figures correspond to eqs.(23) and (24) with $\bar{g}_{{\Lambda^\star}}^2=0.25$,
    $\Gamma_{{\Lambda^\star}}=50$ MeV and $\alpha_{p,n}=\beta_{p,n}$, fixed in the way
    described in the text. The first kink corresponds to the $KN$ threshold 
    and the second around 1.5 GeV to $\sqrt s=m_B+ M_\eta$ for 
    $M_\eta\approx 547$ MeV. 
    The solid line of ${\cal R}e$ part correspond to $Z=-0.5$ and the
    dashed line correspond to $Z=0.15$,
    and the imaginary parts are independent of $Z$ value.
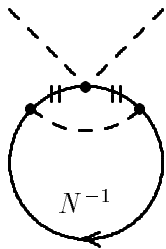
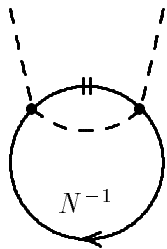
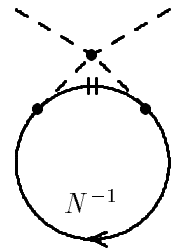
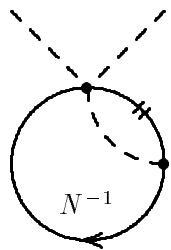
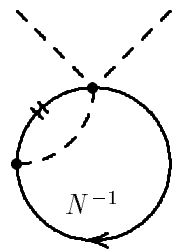
\item {\bf Figure 3:}
    (a): The linear density approximation to the kaon self-energy in medium, $\Pi_K$.
    The square blob represents the off-shell $K^-N$ amplitude
    calculated to $\O (Q^3)$;
    (b)-(f): medium corrections to ${\cal T}^{KN}$ of fig.(a) with the free nucleon
    propagator indicated by a double slash replaced by an in-medium one,
    eq.(\ref{prop}). The loop labeled $\rho_N$ represents the in-medium nucleon loop
    proportional to density, $N^{-1}$ the nucleon hole ($n^{-1}$ 
and/or $p^{-1}$),
    the external dotted line stands for the $K^-$ and the internal dotted line 
    for the pseudoscalar octet $\pi$, $\eta$, $K$.
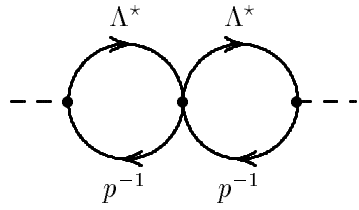
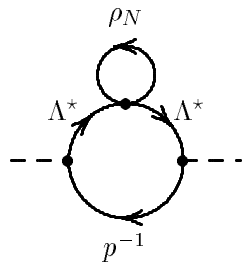
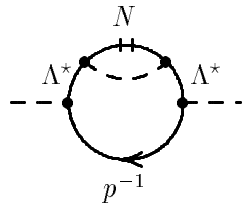
\item {\bf Figure 4:}
    Two-loop diagrams involving $\Lambda^\star$ contributing
    to the kaon self-energy. The diagrams (a)
    and (b) involve four-Fermi interactions, while the
    diagram (c) does not involve four-Fermi interactions
    and hence can be unambiguously determined by on-shell parameters. Here the
    internal dotted line represents the kaon.
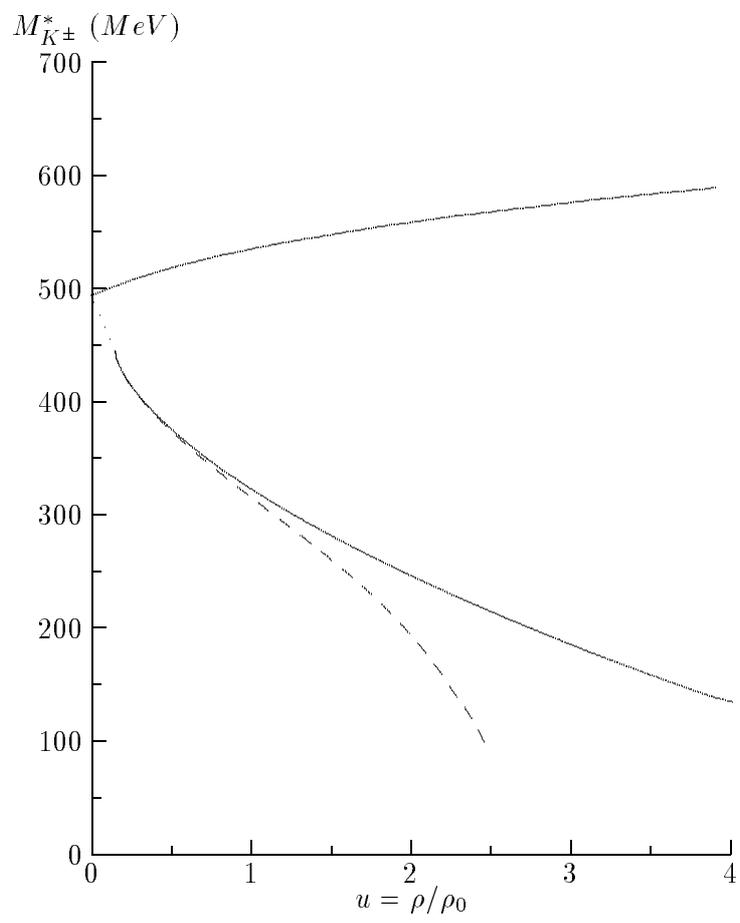
\item {\bf Figure 5:} 
    Plot of the $K^\pm$ effective potential in nuclear matter ($x=0.5$).
    The upper solid line corresponds to $K^+$ ``effective mass" without 
    the {\it BR scaling},
    and the lower solid (dashed) line to $K^-$ ``effective mass" without
    (with) the {\it BR scaling}.
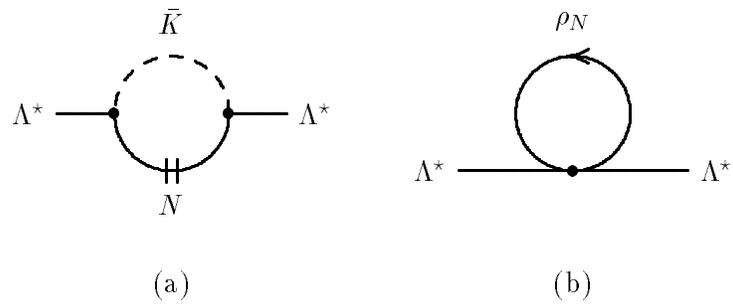
\item {\bf Figure 6:} One-loop diagrams contributing to the $\Lambda^\star$
    mass shift in dense medium. The diagram (a) involves the intermediate
 states of $K^- p$ and $\bar{K}^0 n$ and (b) the four-Fermi interaction
 $C^S_{\Lambda^\star}$ with both protons and neutrons.
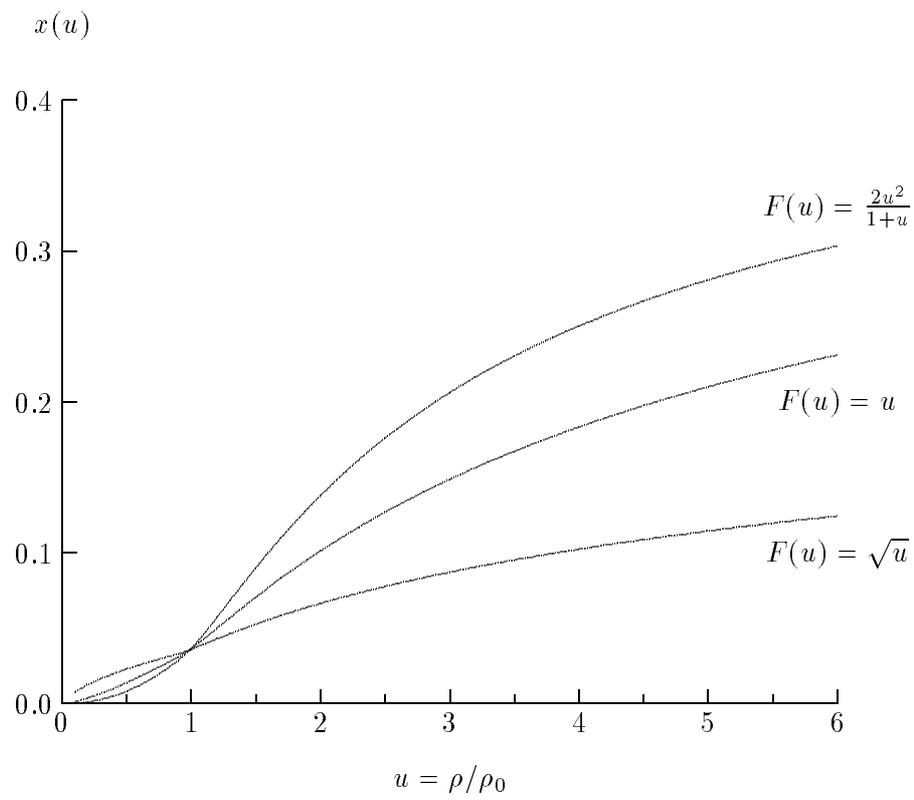
\item {\bf Figure 7:}
    Plot of the proton fraction $x(u)$ prior to kaon condensation
for different forms of $F(u)$.
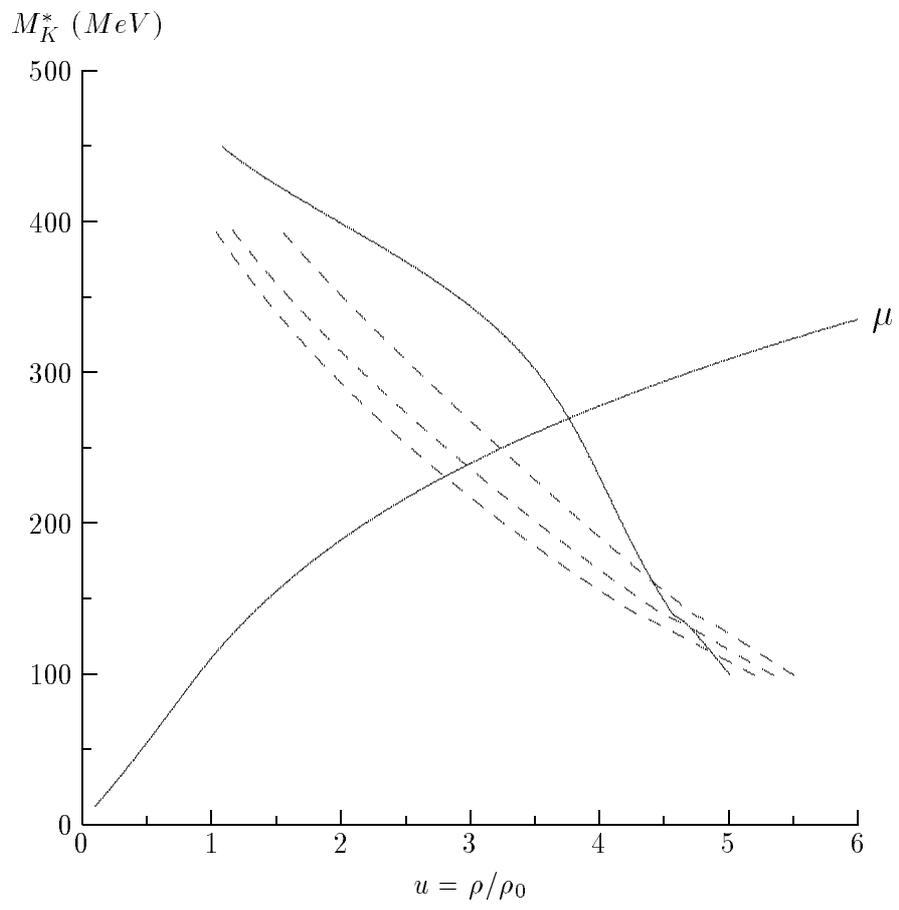
\item {\bf Figure 8:}
    Plot of the quantity $M^\star_K$ obtained from
    the dispersion formula $D^{-1}(\mu,u)$ $=0$ vs. the chemical
    potential $\mu$ prior to kaon condensation for $g_{\Lambda^\star}^2=0.25$
    and $F(u)=2 u^2/(1+u)$. 
    The solid line corresponds to the linear density approximation and
    the dashed lines to the in-medium two-loop results for
    $(C_{\Lambda^\star}^S-C_{\Lambda^\star}^T) f^2 =10$ and
    $C_{\Lambda^\star}^S  f^2 = 10, 5, 0$ respectively from the left.
    The point at which  the chemical potential $\mu$ 
    intersects $M_K^\star$ corresponds to the critical point.
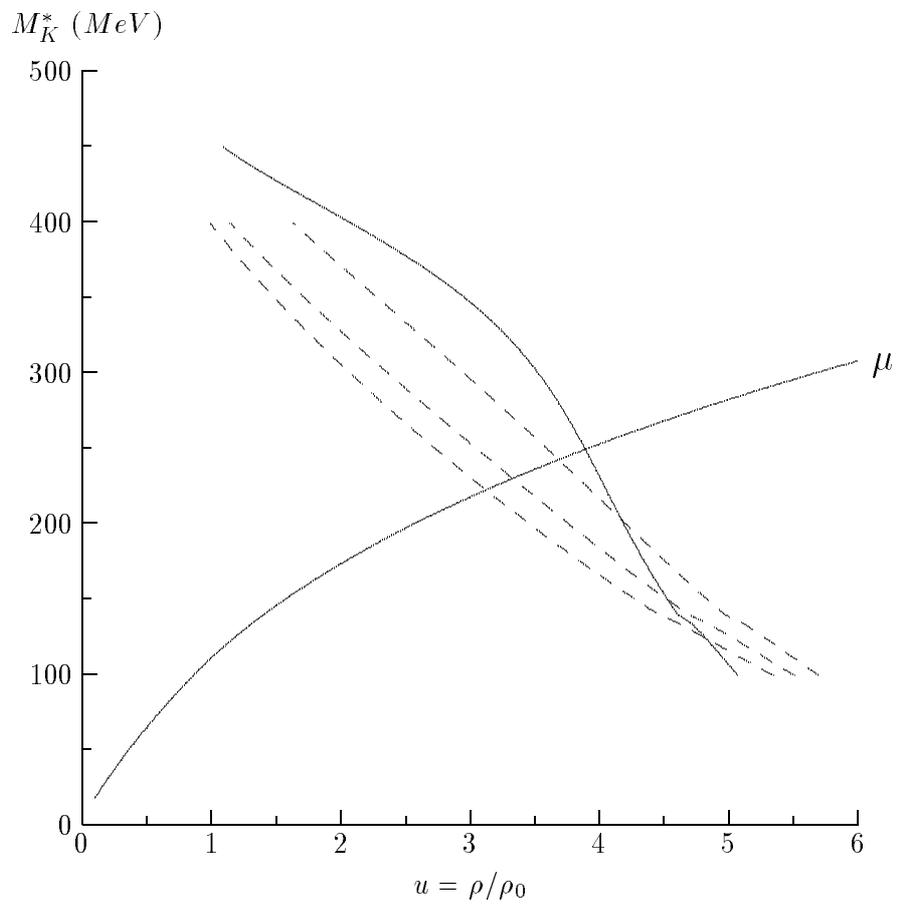
\item {\bf Figure 9:}
    The same as Figure 8 for $F(u)=u$.
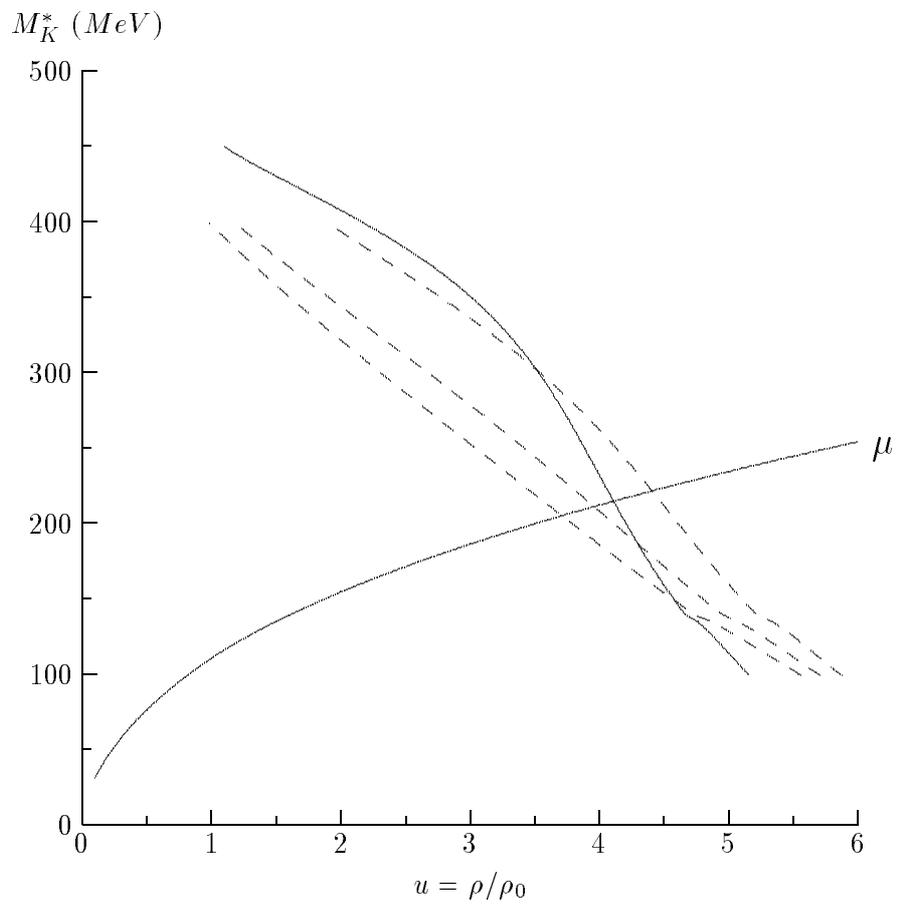
\item {\bf Figure 10:}
    The same as Figure 8 for $F(u)=\sqrt u$.
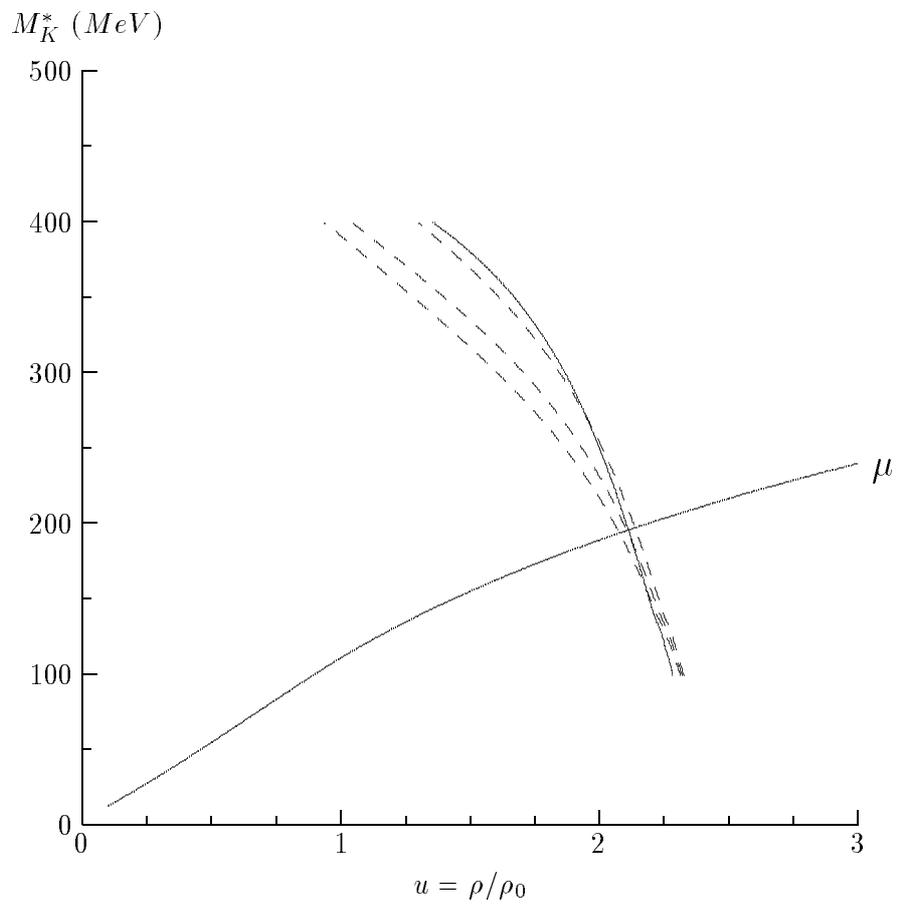
\item {\bf Figure 11:}
    The same as Figure 8 with the {\it BR scaling} for $F(u)=2 u^2/(1+u)$.
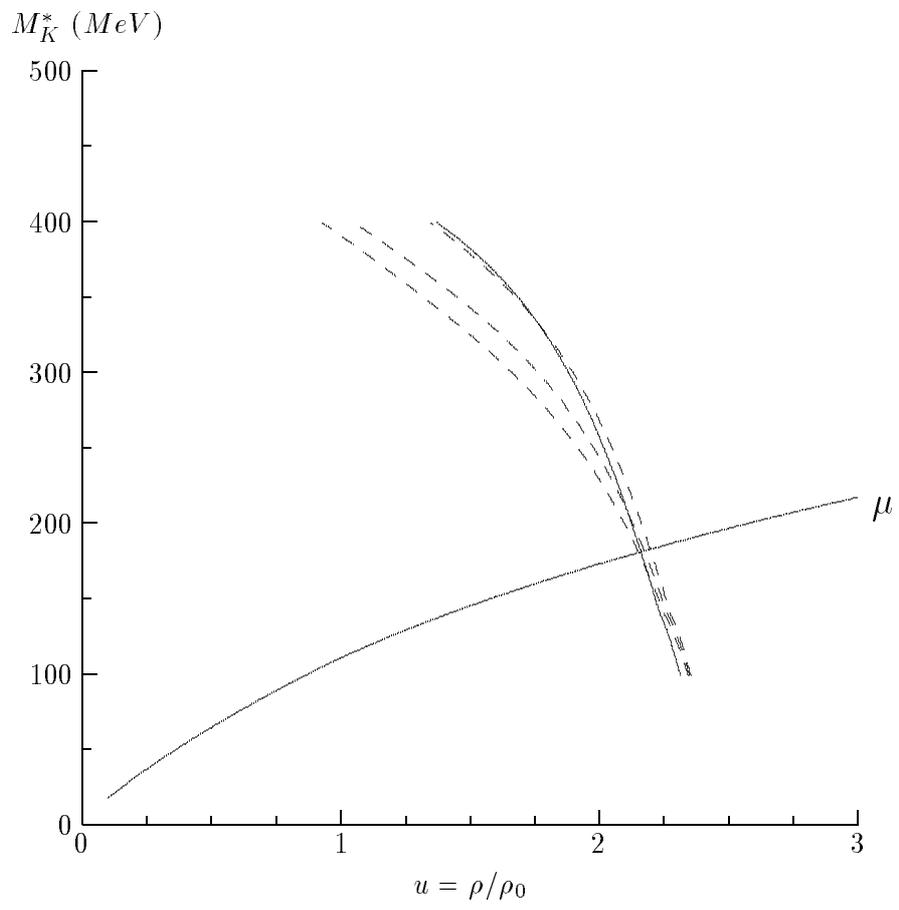
\item {\bf Figure 12:}
    The same as Figure 8 with the {\it BR scaling} for $F(u)= u$.
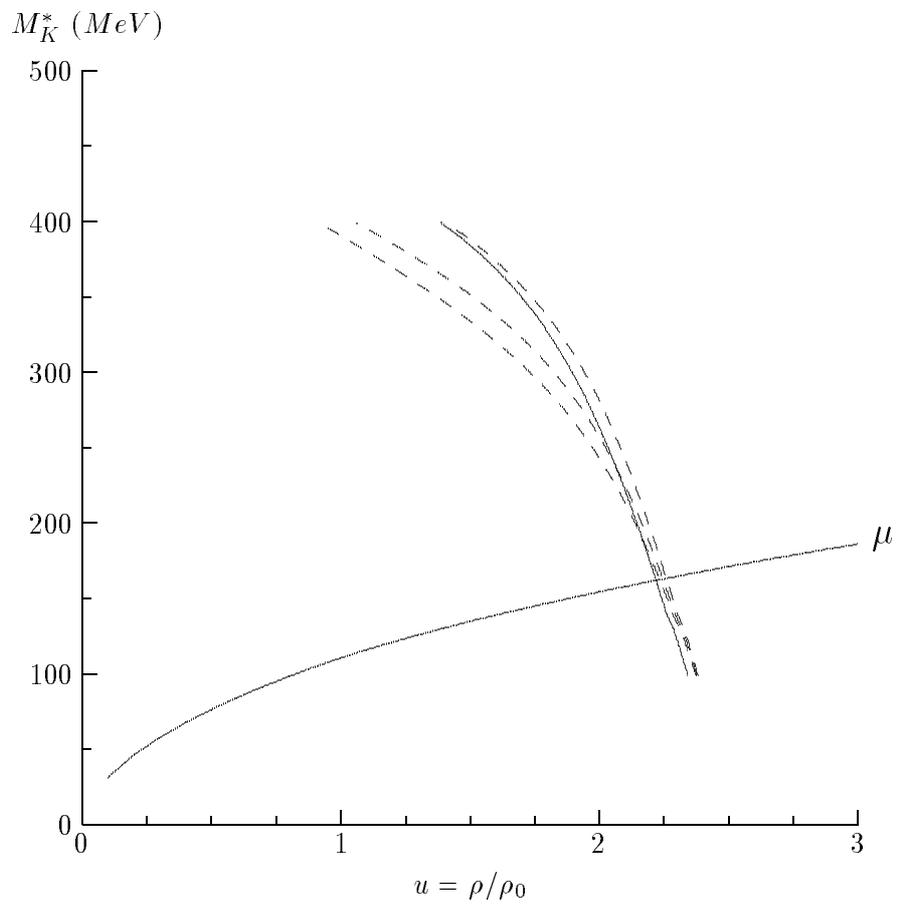
\item {\bf Figure 13:}
    The same as Figure 8 with the {\it BR scaling} for $F(u)=\sqrt u$.
\end{itemize}

\end{document}